\documentclass[runningheads,11pt]{llncs}
\usepackage[top=2.54cm, bottom=2.54cm, left=2.54cm, right=2.54cm]{geometry}
\usepackage[T1]{fontenc}
\usepackage{amsfonts,amsmath,amssymb}
\usepackage{mathtools}
\usepackage{mathrsfs}
\usepackage{upgreek}
\usepackage{cancel}
\usepackage[hidelinks]{hyperref}
\usepackage[sectionbib]{natbib} 
\setcitestyle{aysep={}} 
\usepackage{doi}
\usepackage{xcolor}
\usepackage{graphicx}
\usepackage{algorithm}
\usepackage{algorithmic}
\usepackage{fancyhdr} 
\usepackage{enumitem}

\hypersetup{
  colorlinks=true,
  linkcolor=black,
  urlcolor=blue,
  citecolor=black,
  filecolor=black
}

\newcommand{\bvt}{{\mathbf{t}}}
\newcommand{\bvf}{{\mathbf{f}}}
\newcommand{\bvg}{{\mathbf{g}}}
\newcommand{\bvm}{{\mathbf{m}}}
\newcommand{\bvh}{{\mathbf{h}}}
\newcommand{\bA}{{\mathbf{A}}}
\newcommand{\bB}{{\mathbf{B}}}
\newcommand{\bM}{{\mathbf{M}}}
\newcommand{\bL}{{\mathbf{L}}}

\newcommand{\bH}{{\mathbf{H}}}
\newcommand{\bQ}{{\mathbf{Q}}}
\newcommand{\bP}{{\mathbf{P}}}
\newcommand{\bR}{{\mathbf{R}}}
\newcommand{\bJ}{{\mathbf{J}}}
\newcommand{\bE}{{\mathbf{\Sigma}}}
\newcommand{\bK}{{\mathbf{K}}}
\newcommand{\bI}{{\mathbf{I}}}
\newcommand{\bO}{{\mathbf{O}}}

\newcommand{\tX}{{\mathrm{X}}}
\newcommand{\tx}{{\mathrm{x}}}

\newcommand{\tY}{{\mathrm{Y}}}
\newcommand{\tE}{{\mathrm{E}}}
\newcommand{\tU}{{\mathrm{U}}}
\newcommand{\tu}{{\mathrm{u}}}
\newcommand{\tV}{{\mathrm{V}}}
\newcommand{\tv}{{\mathrm{v}}}
\newcommand{\tW}{{\mathrm{W}}}
\newcommand{\tw}{{\mathrm{w}}}
\newcommand{\tZ}{{\mathrm{Z}}}

\newcommand{\ty}{{\mathrm{y}}}
\newcommand{\tz}{{\mathrm{z}}}

\newcommand{\tb}{{\mathrm{b}}}

\newcommand{\tth}{{\uptheta}}

\newcommand{\Ttu}{{\widetilde \tu}}
\newcommand{\Ttv}{{\widetilde \tv}}

\newcommand{\Ttz}{{\widetilde \tz}}
\newcommand{\Tty}{{\widetilde \ty}}
\newcommand{\TtZ}{{\widetilde \tZ}}
\renewcommand{\qed}{\hfill\blacksquare}

\graphicspath{{figures/}}

\begin{document}

\title{Exact nonlinear state estimation}

\author{{Hristo G. Chipilski}\thanks{\footnotesize This Work has been accepted to the \textit{Journal of the Atmospheric Sciences}. The oﬃcial Version
of Record (VoR) is hosted on the AMS journal server and can be accessed
at \url{https://doi.org/10.1175/JAS-D-24-0171.1}.} 
\href{https://orcid.org/0000-0003-3287-0038}{\includegraphics[scale=0.08]{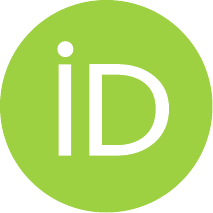}}}

\institute{Department of Scientific Computing, Florida State University, Tallahassee, Florida \\ Advanced Study Program, National Center for Atmospheric Research, Boulder, Colorado \\ \email{hchipilski@fsu.edu}}
\maketitle

\pagestyle{fancy}
\fancyhf{}
\fancyhead[L]{\textit{Exact nonlinear state estimation}}
\fancyhead[R]{\textit{H. G. Chipilski}}
\fancyfoot[C]{\thepage}

\thispagestyle{plain}

\begin{abstract}
The majority of data assimilation (DA) methods in the geosciences are based on Gaussian assumptions. While these assumptions facilitate efficient algorithms, they cause analysis biases and subsequent forecast degradations. Non-parametric, particle-based DA algorithms have superior accuracy, but their application to high-dimensional models still poses operational challenges. Drawing inspiration from recent advances in the field of generative artificial intelligence (AI), this article introduces a new nonlinear estimation theory which attempts to bridge the existing gap in DA methodology. Specifically, a Conjugate Transform Filter (\texttt{CTF}) is derived and shown to generalize the celebrated Kalman filter to arbitrarily non-Gaussian distributions. The new filter has several desirable properties, such as its ability to preserve statistical relationships in the prior state and convergence to highly accurate observations. An ensemble approximation of the new theory (\texttt{ECTF}) is also presented and validated using idealized statistical experiments that feature bounded quantities with non-Gaussian distributions, a prevalent challenge in Earth system models. Results from these experiments indicate that the greatest benefits from \texttt{ECTF} occur when observation errors are small relative to the forecast uncertainty and when state variables exhibit strong nonlinear dependencies. Ultimately, the new filtering theory offers exciting avenues for improving conventional DA algorithms through their principled integration with AI techniques.

\keywords{nonlinear state-space models \and tractable non-Gaussian filtering distributions \and Ensemble Conjugate Transform Filter (\texttt{ECTF})}
\end{abstract}

\section{Introduction} \label{sec:intro}

The field of DA has experienced tremendous expansion in recent decades due to advances in Earth system modeling and observations, as well as steady growth in the available computational resources. However, the restrictive linear-Gaussian assumptions made in standard DA algorithms lead to systematic errors in initial conditions and subsequent forecasts \citep[e.g., see][]{poterjoy_2022}, presenting limitations to future progress. 

Methods for accurately solving the DA problem exist and are based on non-parametric Monte Carlo techniques. A popular example in this category is the Particle Filter \citep[\texttt{PF};][]{gordon_et_al_1993}. \texttt{PF}s have undergone significant developments in recent decades \citep{vanLeeuwen_2009,vanLeeuwen_et_al_2019}, including their successful application to a variety of high-dimensional geophysical models \citep{todter_et_al_2016,poterjoy_et_al_2017,rojahn_et_al_2023}. Simultaneously, there have been efforts to relax the Gaussian approximations in standard DA algorithms, with prominent examples being the lognormal extensions of variational and Kalman filter (\texttt{KF}) methods \citep{fletcher_zupanski_2006a,fletcher_zupanski_2007,fletcher_2010,fletcher_et_al_2023} and the family of two-step ensemble filters \citep{anderson_2003,anderson_2010,anderson_2022,grooms_2022}. Nevertheless, recent findings have shown that more work is still needed to fully exploit the potential of nonlinear DA. For examples, \citet{poterjoy_2022} and \citet{rojahn_et_al_2023} found that \texttt{PF}s perform either similarly or marginally better in comparison with operational DA methods. As far as non-Gaussian extensions of standard DA algorithms are concerned, distributional assumptions are either still restrictive or only made in observation space.  

In an attempt to bridge the aforementioned methodological gap, this study develops a new nonlinear filtering theory which generalizes the popular Kalman filter \citep{kalman_1960}, the basis for many geophysical DA algorithms. Unlike other non-Gaussian extensions, the new filtering approach boasts the flexibility to incorporate arbitrary analysis distributions, bringing it closer to \texttt{PF}s and other non-parametric Monte Carlo techniques. The mathematical developments presented herein take advantage of recent advances in probabilistic deep learning and align with the ongoing surge of studies at the intersection of DA and AI \citep[for a review, see][]{bocquet_2023}.

This paper is organized as follows. In Section \ref{sec:prob_view_estimation}, a probabilistic view of estimation theory is first outlined focusing on the inability to obtain closed-form solutions with the standard nonlinear state-space model. Section \ref{sec:CTF} resolves this obstacle by defining an alternative state-space model which leads to exact nonlinear filtering recursions. Section \ref{sec:ECTF} formulates an ensemble approximation of the resulting Conjugate Transform Filter (\texttt{ECTF}) whose Bayesian consistency and analysis benefits are further illustrated through the idealized statistical experiments in Section \ref{sec:exps}. A summary of all important theoretical and experimental results follows in Section \ref{sec:conclusions}, together with a brief discussion of future research directions. Finally, all mathematical proofs and additional discussions are given in a supplementary appendix.

\section{Probabilistic view of estimation theory} \label{sec:prob_view_estimation}

Estimation theory \citep{jazwinski_1970} offers natural mathematical tools to derive contemporary DA algorithms \citep{cohn_1997}. State-space models (SSMs) are a key element in this theory and can be defined as a pair of two stochastic processes, $\{ \tX_k \}_{k \in \mathbb{Z}_{\geq 0}}$ and $\{ \tY_k \}_{k \in \mathbb{Z}_{>0}}$, which describe the temporal evolution of the dynamical system and the generation of new observations within this system. In particular, $\forall k \in \mathbb{Z}_{>0}$,

\vspace*{-2mm}
\begin{subequations}
\label{eq:SSM}
\begin{align}
    &\tX_k = \bvf_k \left(\tX_{k-1}, \tE^m_{k-1} \right) \label{eq:SSM_X}, \\
    &\tY_k = \bvg_{k} \left(\tX_k, \tE^o_k \right) \label{eq:SSM_Y},
\end{align}
\end{subequations}
\vspace*{-2mm}

\noindent where $\bvf_k: \mathbb{R}^{N_x} \times \mathbb{R}^{N_x} \to \mathbb{R}^{N_x}$ and $\bvg_k: \mathbb{R}^{N_x} \times \mathbb{R}^{N_y} \to \mathbb{R}^{N_y}$. The state and observations are additionally corrupted by the error processes $\{ \tE^m_k \}_{k \in \mathbb{Z}_{\geq 0}}$ and $\{ \tE^o_k \}_{k \in \mathbb{Z}_{>0}}$.

Assuming that all random variables in the SSM admit probability density functions (pdfs) with respect to the standard Lebesgue measure, the estimation problem can be defined more compactly in probabilistic terms. Using the notation $\{ m:n \} \coloneqq \{ m, m+1, ..., n-1, n \}$, the primary object of interest is the conditional density $p(\tx_{0:T} | \ty_{1:T})$, where $T \in \mathbb{Z}_{>0}$ represents the time window over which observations are assimilated. Different marginals of this conditional pdf correspond to different estimation tasks. For example, in filtering, which is the subject of this paper, the goal is to approximate the conditional pdf $p(\tx_k | \ty_{1:k})\ \forall k \in \{1:T\}$.  

\subsection{Graphical models}

\begin{figure}[!ht]
  \centering
  \includegraphics[width=0.75\textwidth]{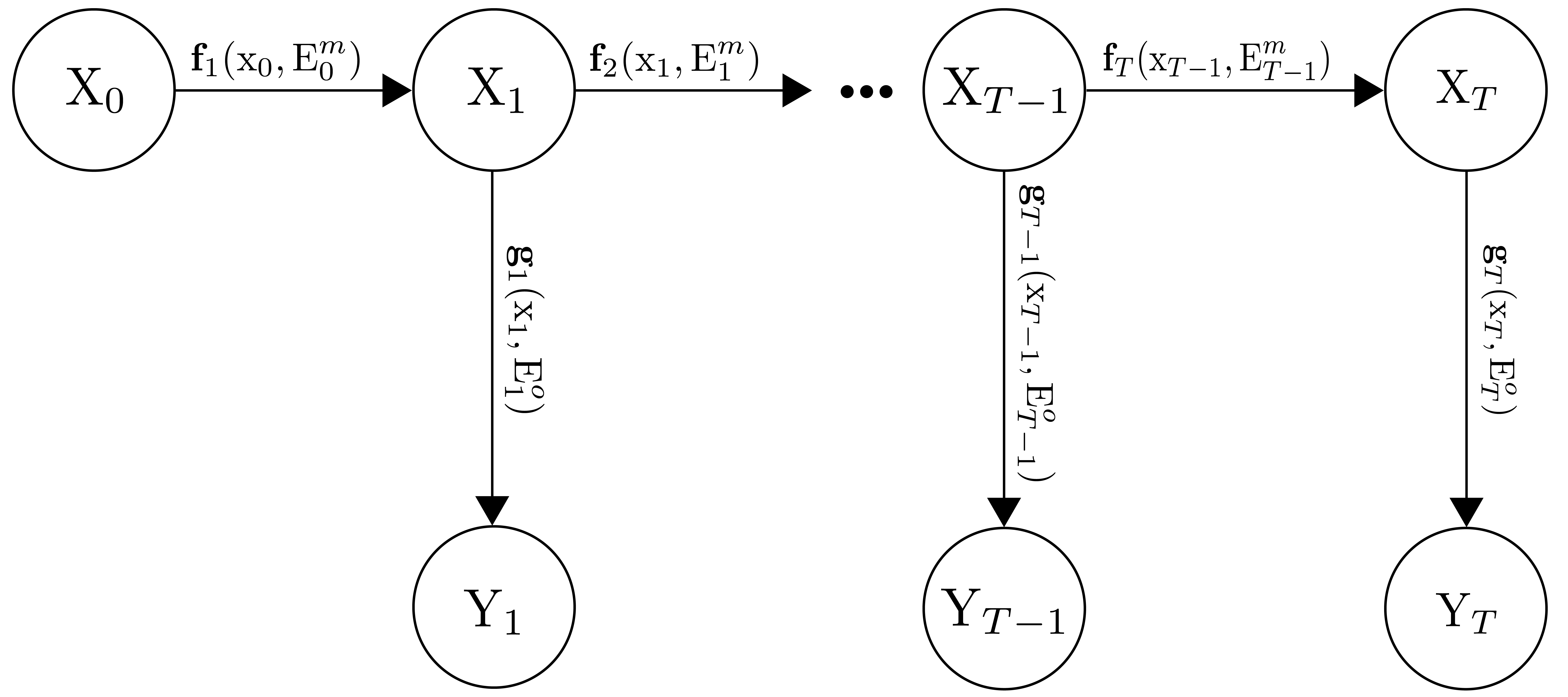}
  \caption{A probabilistic representation of the state estimation problem. The general state-space model (\ref{eq:SSM}) is integrated into a hidden Markov model, with horizontal and vertical arrows corresponding to the state transition dynamics and the observation generation process, respectively.}
  \label{fig:HMM}
\end{figure}

This probabilistic nature of estimation theory can be conveniently captured through the lens of directed acyclic graphs (DAGs). The main purpose of such graphical models is to encode the conditional dependencies between a collection of random variables. The hidden Markov model (HMM) depicted in Fig.~\ref{fig:HMM} is one example of a DAG and is a natural choice to represent the estimation problem. The random variables in this case consist of the state and observation stochastic processes; i.e., $\{ \tX_0, \tX_1, ..., \tX_T, \tY_1, ..., \tY_T \}$. The dependency structure of the HMM allows for the decomposition of the conditional pdf related the general estimation problem; that is,

\begin{equation} \label{eq:conditional_pdf}
   p\left( \tx_{0:T} | \ty_{1:T} \right) \underset{\tx_{0:T}}{\propto} p\left( \tx_{0:T},\ty_{1:T} \right) = p\left( \tx_0 \right) \prod_{k=1}^{T} p(\tx_k|\tx_{k-1}) p(\ty_k|\tx_k).
\end{equation}
\vspace*{0.5mm}

The two conditional pdfs appearing in the last expression above are referred to as the state transition and emission densities\footnote{Emission densities and likelihoods are often used interchangeably in the literature.} and reveal two important properties of the HMM: (i) the dynamical evolution depends only on the previous state (Markov property) and (ii) observations are generated from the current system state. These properties are easily visualized with the directed arrows in Fig.~\ref{fig:HMM}. 

In the context of estimation theory, the factorization in (\ref{eq:conditional_pdf}) provides a natural mechanism to incorporate SSMs into the HMM: the two conditional densities in (\ref{eq:conditional_pdf}) can be calculated from the state transition and observation models in (\ref{eq:SSM}). The only required technical modification is to replace the random variable $\tX$ on the right-hand side of the SSM with its (deterministic) realization $\tx$ (\textit{cf}.~Appendix \ref{appendix:notation} for more details on the notation used throughout this paper).

\subsection{The standard SSM and its Gaussian limitations}

Equation (\ref{eq:SSM}) represents the most general form of the SSM for discrete dynamical systems. To enable further mathematical analysis, some approximations are needed. It is standard practice to assume that $\{ \tE^m_k \}_{k \in \mathbb{Z}_{\geq 0}}$ and $\{ \tE^o_k \}_{k \in \mathbb{Z}_{>0}}$ consist of independent and identically distributed random variables, which are additive and Gaussian distributed. This leads to the following simplified SSM:

\vspace*{-3mm}
\begin{subequations}
\label{eq:SSM-aG}
\begin{align}
    &\tX_k = \bvm_k \left( \tx_{k-1} \right) + \tE^m_{k-1} &\text{with}\ &\tE_{k-1}^m \sim \mathcal{N}(0,\bQ_{k-1}), \label{eq:SSM_X-aG} \\
    &\tY_k = \bvh_k \left(\tx_k \right) + \tE^o_k &\text{with}\ &\tE_k^o \sim \mathcal{N}(0,\bR_k) \label{eq:SSM_Y-aG}.
\end{align}
\end{subequations}
\vspace*{-3mm}

\noindent The appealing property of this additive-Gaussian SSM is that the noise terms are separated from the potentially nonlinear deterministic functions $\bvm_k$ and $\bvh_k$, allowing for a clearer interpretation of its meaning: $\bvm_k$ represents the discretized numerical model, whereas $\bvh_k$ is the observation operator mapping state variables to observation space $\mathcal{Y} \subset \mathbb{R}^{N_y}$. 

To calculate $p(\tx_k|\tx_{k-1})$ and $p(\ty_k|\tx_k)$, the change of variables theorem can be used. Let $\tW_1$ and $\tW_2$ be two random variables connected through the diffeomorphism\footnote{A diffeomorphism $\bvt$ is an invertible function with the additional requirements that $\bvt$ and $\bvt^{-1}$ are both differentiable.} $\bvt: \mathbb{R}^N \to \mathbb{R}^N$ such that $\tW_2 = \bvt(\tW_1)$. Assuming that $\tW_1$ admits the pdf $p_{\scriptscriptstyle \tW_1}(\tw_1)$, the pdf of $\tW_2$ can be written as

\begin{equation} \label{eq:change_of_variables}
   p_{\scriptscriptstyle \tW_2}(\tw_2) = p_{\scriptscriptstyle \tW_1} \left( \bvt^{-1}(\tw_2) \right) | det\ \bJ_{\bvt^{-1}}(\tw_2) |,
\end{equation}
\vspace*{-2mm}

\noindent where $\bJ_{\bvt^{-1}}(\tw_2)$ is the Jacobian matrix of the inverse transformation $\bvt^{-1}$. Taking the determinant and absolute value of this Jacobian matrix accounts for the volume changes in the pdf of $\tW_2$ induced by $\bvt$. Applying this result to the additive-Gaussian SSM (\ref{eq:SSM-aG}),

\vspace*{-3mm}
\begin{subequations}
\label{eq:cond_pdfs-aG_SSM}
\begin{align}
    &p(\tx_k|\tx_{k-1}) = \phi \left[\tx_k; \bvm_k \left( \tx_{k-1} \right), \bQ_{k-1}  \right], \label{eq:cond_pdfs-aG_SSM-state} \\
    &p(\ty_k|\tx_k) = \phi \left[\ty_k; \bvh_k \left(\tx_k \right), \bR_k  \right] \label{eq:cond_pdfs-aG_SSM-ob},
\end{align}
\end{subequations}
\vspace*{-3mm}

\noindent where $\phi \left[ \cdot; \upmu, \bE \right]$ denotes a Gaussian pdf with a mean $\upmu$ and covariance $\bE$.

The Gaussian structure of the state transition and emission pdfs is quite intuitive given that the deterministic shift vectors $\bvm_k \left( \tx_{k-1} \right)$ and $\bvh_k \left(\tx_k \right)$ are added to the zero-centered Gaussian noise terms $\tE^m_{k-1}$ and $\tE^o_k$. However, the discussion in Section \ref{sec:intro} already hinted that such Gaussian assumptions are not appropriate for many geophysical applications. The second more serious problem is that the nonlinearities in $\bvm_k$ and $\bvh_k$ make it impossible to obtain closed-form (exact) solutions to the general estimation problem captured by $p\left( \tx_{0:T} | \ty_{1:T} \right)$ \citep[e.g., see][]{calvello_et_al_2023}. Accurate estimation in this nonlinear setting can be achieved only through expensive DA methods like the \texttt{PF}. The next section resolves this obstacle by formulating an alternative SSM, which not only incorporates highly nonlinear functions, but also produces tractable filtering distributions with arbitrary structure.

\section{Conjugate Transform Filter}
\label{sec:CTF}
As discussed earlier, the Gaussian restrictions of the standard SSM (\ref{eq:SSM-aG}) have given rise to some alternative nonlinear formulations such as the lognormal SSM of \citet[][Section 5.3]{cohn_1997}, which is suitable for positive definite variables with multiplicative errors. Indeed, this probabilistic model serves as the basis for all lognormal variants of standard DA algorithms mentioned in Section \ref{sec:intro}. However, one limitation of such approaches is the limited representation power of the multivariate lognormal distribution, especially in the presence of multimodal statistics.

\subsection{Deep probabilistic modeling}

Alternatively, recent advances in deep learning have opened the doors for the construction of more complex distributions by composing several diffeomorphisms in a chain $\bvt \coloneqq \bvt_n \circ \bvt_{n-1} \circ ... \circ \bvt_2 \circ \bvt_1$, each represented by a set of learnable parameters. Since the composite transformation $\bvt$ is also a diffeomorphism \citep{papamakarios_et_al_2021}, the change-of-variables theorem (\ref{eq:change_of_variables}) still applies. In fact, is possible to express the transformed pdf in terms of the individual functions in the composition by noting that

\vspace*{-3mm}
\begin{align}
    &\bvt^{-1} = \left( \bvt_n \circ \bvt_{n-1} \circ ... \circ \bvt_2 \circ \bvt_1 \right)^{-1} = \bvt_1^{-1} \circ \bvt_2^{-1} \circ ... \circ \bvt_{n-1}^{-1} \circ \bvt_{n}^{-1}, \\
    &det\ \bJ_{\bvt^{-1}} = det\ \bJ_{\bvt_1^{-1} \circ ... \circ \bvt_n^{-1}} = \prod_{i=1}^{n} det\ \bJ_{\bvt_i^{-1}},
\end{align}
\vspace*{-1mm}

\noindent where it is understood that $\bJ_{\bvt_n^{-1}} \coloneqq \bJ_{\bvt_n^{-1}}(\tw_2)$, $\bJ_{\bvt_{n-1}^{-1}} \coloneqq \bJ_{\bvt_{n-1}^{-1}}(\bvt_n^{-1}(\tw_2))$, etc. 

The resulting architectures are known as Invertible Neural Networks (INNs) and have already been used to define new SSMs. In their differentiable \texttt{PF}s, \citet{chen_et_al_2021} modify the state transition equation (\ref{eq:SSM_X}) to derive more flexible proposal distributions that steer particles closer to observations. \citet{de_bezenac_et_al_2020} takes an alternative approach by introducing nonlinearities in the observation model (\ref{eq:SSM_Y}) to enhance the analysis of multivariate time series. An important feature of their Normalizing Kalman Filter (\texttt{NKF}) is that it offers exact solutions to the filtering problem. However, one notable limitation is that the posterior pdfs $p(\tx_k | \ty_{1:k})$ remain Gaussian.

\subsection{An alternative nonlinear SSM}

Addressing the fully non-Gaussian estimation problem requires modifications to both the state and observation equations. Inspired by the previous developments, this work proposes the following new SSM: 

\vspace*{-3mm}
\begin{subequations}
\label{eq:new_SSM}
\begin{align}
  & \tX_k = \bvf_k \left( \bM_k \widetilde{\tx}_{k-1} + \tE_{k-1}^m; \, \tth_k^f \right) & \text{with}\ & \tE_{k-1}^m \sim \mathcal{N}(0,\bQ_{k-1}) \label{eq:new_SSM_X},  \\
  & \tY_k = \bvg_k \left(\bH_k \widetilde{\tx}_k + \tE_k^o; \, \tth_k^g \right) & \text{with}\ & \tE_k^o \sim \mathcal{N}(0,\bR_k) \label{eq:new_SSM_Y},
\end{align}
\end{subequations}
\vspace*{-3mm}

\noindent with tilde symbols denoting variables in the latent (reference) Gaussian space; e.g., $\widetilde{\tx}_k \coloneqq \bvf_k^{-1} \left( \tx_k \right)$. This SSM can be derived by making two approximations to the general SSM in (\ref{eq:SSM}):

\begin{itemize}[leftmargin=*]
  \setlength\itemsep{0.05mm}
  \item The state and observation functions $\bvf_k$ and $\bvg_k$ are two nonlinear diffeomorphisms parameterized by the vectors $\tth^f_k$ and $\tth^g_k$.
  \item The input to $\bvf_k$ and $\bvg_k$ comes from an auxiliary linear-Gaussian SSM with a model propagator $\bM_k$ and an observation operator $\bH_k$.
\end{itemize}

\noindent The introduction of parameters in $\bvf_k$ and $\bvg_k$ is intended to keep the discussion as general as possible and relate the new SSM to INNs. While these functions can be made arbitrarily complex by stacking several learnable diffeomorphisms together, it is also possible to use much simpler (non-parametric) functions. In fact, replacing $\bvf_k$ and $\bvg_k$ with exponentials recovers the lognormal observation model of \citet{cohn_1997}.

The new SSM (\ref{eq:new_SSM}) has a clear and intuitive interpretation: the functions $\bvf_k$ and $\bvg_k$ apply nonlinear corrections to a linear-Gaussian dynamical system. Having learnable parameters as part of these nonlinear corrections reflects our partial knowledge of the system dynamics. It is also worth mentioning that the structure of the proposed SSM resembles a classical neural network where affine transformations of the input vector are augmented with nonlinear activation functions.

In contrast to the additive-Gaussian formulation, the new SSM induces the following non-Gaussian state transition and emissions densities,

\vspace*{-3mm}
\begin{subequations}
\label{eq:cond_pdfs-new_SSM}
\begin{align}
    &p(\tx_k|\tx_{k-1}) = \bvf_{\tth_k \, \sharp} \, \phi \left[\tx_k; \bM_k \widetilde \tx_{k-1}, \bQ_{k-1}  \right], \label{eq:cond_pdfs-new_SSM-state} \\
    &p(\ty_k|\tx_k) = \bvg_{\tth_k \, \sharp} \, \phi \left[\ty_k; \bH_k \widetilde \tx_k, \bR_k  \right] \label{eq:cond_pdfs-new_SSM-ob},
\end{align}
\end{subequations}
\vspace*{-3mm}

\noindent The sharp symbol ($\sharp$) above indicates that $p(\tx_k|\tx_{k-1})$ and $p(\ty_k|\tx_k)$ are obtained by applying the change of variables theorem (\ref{eq:change_of_variables}) to the two Gaussian pdfs on the right-hand side; in other words, these Gaussian pdfs are \textit{pushed forward}\footnote{The notions of pushing forward and pulling back are typically associated with probability measures (rather than pdfs), but this is common abuse of notation in the data sciences.} by the nonlinear functions $\bvf_{\tth_k}$ and $\bvg_{\tth_k}$. Note once again that the $\tth_k$ subscripts in the state and observation functions emphasize their potential dependence on parameters that can be further estimated.

\subsection{Filtering distributions}

Having determined the forms of $p(\tx_k|\tx_{k-1})$ and $p(\ty_k|\tx_k)$, the only remaining piece needed to solve the filtering problem is to specify that the initial state distribution is given by 

\vspace*{-2.5mm}
\begin{equation}
    p(\tx_0) = \bvf_{\tth_0 \, \sharp} \, \phi \left[\tx_0 ;\upmu_0,\bE_0 \right]. \label{eq:initial_pdf}
\end{equation} 
\vspace*{-3mm}

\noindent To calculate the posterior $p(\tx_k | \ty_{1:k})\ \forall k \in \{1:T\}$, recall that all filtering algorithms are sequential in nature and can be expressed recursively as the alternation of prediction and update (analysis) steps \citep[e.g., see eq.~10 in][]{de_bezenac_et_al_2020}; that is,

\vspace*{-3mm}
\begin{align}
    &p(\tx_k | \ty_{1:k-1}) = \int p(\tx_k | \tx_{k-1}) p(\tx_{k-1}|\ty_{1:k-1}) d\tx_{k-1} &\text{(prediction)}, \label{eq:filtering_recursions-prediction} \\
    &p(\tx_k | \ty_{1:k}) = \frac{p(\tx_k|\ty_{1:k-1})p(\ty_k|\tx_k)}{\int p(\tx_k|\ty_{1:k-1})p(\ty_k|\tx_k) d\tx_k} &\text{(update)}\label{eq:filtering_recursions-update}.
\end{align}
\vspace*{1mm}

\noindent Having defined the state transition pdf $p(\tx_k | \tx_{k-1})$, the Chapman-Kolmogorov equation (\ref{eq:filtering_recursions-prediction}) propagates the posterior $p(\tx_{k-1}|\ty_{1:k-1})$ at time $k-1$ to the next time $k$. The resulting density $p(\tx_k|\ty_{1:k-1})$ is referred to as the prior, and is updated with (\ref{eq:filtering_recursions-update}) to form the posterior pdf $p(\tx_{k}|\ty_{1:k})$ at time $k$.

It is widely accepted that closed-form solutions to (\ref{eq:filtering_recursions-prediction}) and (\ref{eq:filtering_recursions-update}) are not feasible except for simple SSMs, such as those featuring linear-Gaussian dynamics. Some recent articles supporting this claim include \citet{ait-el-fquih_et_al_2023}, \citet{chen_li_2023} and \citet{look_et_al_2023}. The next two results offer an alternative perspective to this challenge. First, it is shown that the product of two non-Gaussian pdfs obtained via the change of variables theorem (\ref{eq:change_of_variables}) with Gaussian base densities is analytically tractable.

\vspace*{5mm}
\noindent \textbf{\hypertarget{lemma:prod_nonGauss_pdfs}Lemma} (Closed-form products of non-Gaussian densities). \textit{Given the random vectors $\tU$ and $\tV$, let $\bvt_u: \mathbb{R}^{N_u} \ni \widetilde \tU \mapsto \tU \in \mathbb{R}^{N_u}$ and $\bvt_v: \mathbb{R}^{N_v} \ni \widetilde \tV \mapsto \tV \in \mathbb{R}^{N_v}$ be two nonlinear diffeomorphisms inducing the non-Gaussian densities $p(\tu|\tw) = \bvt_{u\, \sharp} \, \phi \left[\tu ;\upmu_u,\bE_u \right]$ and $p(\tv|\tu) = \bvt_{v\, \sharp} \, \phi \left[\tv ;\bA \widetilde \tu + \tb,\bE_v \right]$, where $\tW$ is another random vector, $\bA \in \mathbb{R}^{N_v \times N_u}$ and $\tb \in \mathbb{R}^{N_v}$. Then it holds that}

\vspace*{-3mm}
\begin{align}
    & p(\tv|\tu) p(\tu|\tw) \underset{\tu}{\propto} \bvt_{u\, \sharp} \, \phi \left[\tu ;\upmu_p,\bE_p \right] \label{eq:lemma_i-pdf},\\
    & \bE_p = \left(\bE_u^{-1} + \bA^\top \bE_v^{-1} \bA \right)^{-1} \label{eq:lemma_i-sigma},\\
    & \upmu_p = \bE_p \left[ \bE_u^{-1}\upmu_u + \bA^\top \bE_v^{-1} (\widetilde \tv-\tb) \right] \label{eq:lemma_i-mu}.
\end{align}
\vspace*{-3mm}

\noindent \textit{In addition, if $\tV$ is conditionally independent of $\tW$ given $\tU$,}

\vspace*{-3mm}
\begin{align}
    p(\tv|\tw) = \int p(\tv|& \tu) p(\tu|\tw) d\tu = \bvt_{v\, \sharp} \, \phi \left[\tv ;\upmu_i,\bE_i \right] \label{eq:lemma_ii-pdf},\\
    & \bE_i = \bA \bE_u \bA^\top + \bE_v \label{eq:lemma_ii-sigma},\\
    & \upmu_i = \bA \upmu_u + \tb \label{eq:lemma_ii-mu}.
\end{align}
\vspace*{-3mm}

\noindent Equipped with this lemma, the next theorem proves that all non-Gaussian filtering densities associated with the nonlinear SSM (\ref{eq:new_SSM}) can be expressed in closed form.

\vspace*{5mm}
\noindent \textbf{\hypertarget{thm:CTF}{Theorem}} (Exact nonlinear filtering). \textit{Suppose that the initial state pdf $p(\tx_0)$ is defined according to (\ref{eq:initial_pdf}). Then, for all $k \in \mathbb{Z}_{>0}$, the filtering (posterior) pdf associated with the SSM (\ref{eq:new_SSM}) is given by}

\vspace*{-1mm}
\begin{equation}
    p(\tx_k | \ty_{1:k}) = \bvf_{\tth_k\, \sharp} \, \phi \left[\tx_k ;\upmu_{k|k},\bE_{k|k} \right], \label{eq:CTF_theorem-pdf}
\end{equation} 

\vspace*{1mm}
\noindent \textit{where}

\vspace*{-5mm}
\begin{align} 
    & \upmu_{k|k} = \upmu_{k|k-1} + \bK_k \left( \widetilde \ty_k - \bH_k \upmu_{k|k-1} \right) \label{eq:CTF_theorem-mu},  \\
    & \bE_{k|k} = \left( \bI-\bK_k \bH_k  \right) \bE_{k|k-1} \label{eq:CTF_theorem-sigma}, \\
    & \bK_k = \bE_{k|k-1} \bH_k^\top \left( \bH_k \bE_{k|k-1} \bH_k^\top + \bR_k  \right)^{-1} \label{eq:CTF_theorem-gain}.
\end{align}

\noindent \textit{The latent Gaussian parameters $\upmu_{k|k-1}$ and $\bE_{k|k-1}$ correspond to the prior density}

\vspace*{-1mm}
\begin{equation} \label{eq:CTF_theorem-prior_pdf}
    p(\tx_k | \ty_{1:k-1}) = \bvf_{\tth_k\, \sharp} \, \phi \left[\tx_k ;\upmu_{k|k-1},\bE_{k|k-1} \right]
\end{equation}

\noindent \textit{and are defined as}

\vspace*{-5mm}
\begin{align} 
    & \upmu_{k|k-1} = \bM_k \upmu_{k-1|k-1} \label{eq:CTF_theorem-mu_prior},  \\
    & \bE_{k|k-1} = \bM_k \bE_{k-1|k-1} \bM_k^\top + \bQ_{k-1} \label{eq:CTF_theorem-sigma_prior}.
\end{align}
\vspace*{-1mm}

\subsection{Properties} \label{subsec:CTF_properties}

A valuable feature of the newly derived filtering solutions is that the prior $p(\tx_k | \ty_{1:k-1})$ and posterior $p(\tx_k | \ty_{1:k})$ belong to the same distribution family parameterized by the nonlinear state function $\bvf_{\tth_k}$. In the context of Bayesian inference, the prior is said to be \textit{conjugate} to the likelihood, which inspires the name \textit{Conjugate Transform Filter} (\texttt{CTF}). Selecting different transformations $\bvf_{\tth_k}$ and $\bvg_{\tth_k}$ is equivalent to adaptively choosing different prior-likelihood pairs. On the other hand, previous conjugate filtering algorithms like the \texttt{GIGG-EnKF} of \citet{bishop_2016} or the selection \texttt{EnKF} of \citet{conjard_omre_2020} are restricted to specific distribution classes, which limits their representation power. From a dynamical perspective, the conjugate update has the additional benefit of preserving statistical relationships and physical bounds in the prior state -- an important property which will be demonstrated numerically in Section \ref{subsec:exps_results}.

\begin{figure}[!ht]
  \centering
  \includegraphics[width=0.98\textwidth]{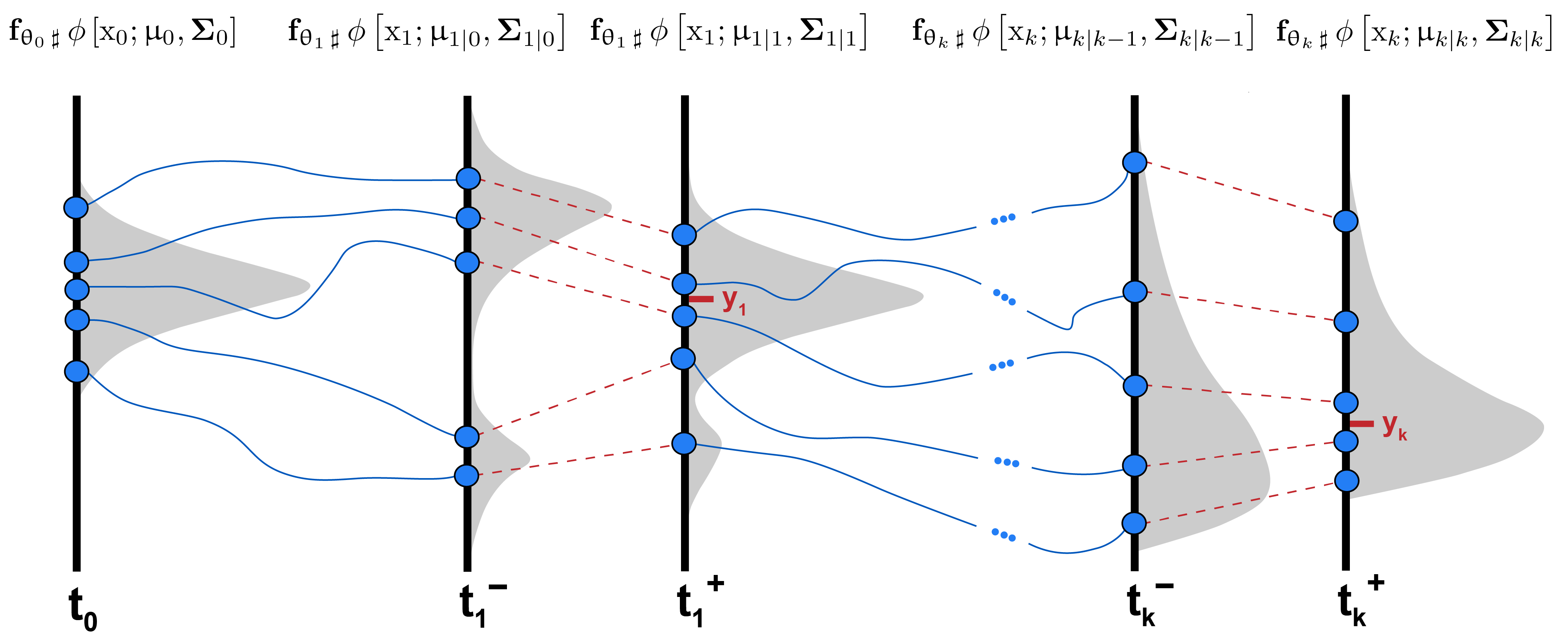}
  \caption{A schematic illustration of the Conjugate Transform Filter (\texttt{CTF}). Starting with a Gaussian distribution (gray shading) at time $t_0$, the state is propagated to time $t_1$ when the first observations $\ty_1$ become available. Owing to the nonlinear dynamics (thin blue curves), the initial state distribution deforms and results in a non-Gaussian prior, which needs to be approximated through a suitable choice of $\bvf_{\tth_1}$. The observations $\ty_1$ are then used by \texttt{CTF} to update the prior, but only the latent Gaussian parameters $\{\upmu_{1|0}, \bE_{1|0}\}$ are modified, whereas the values of $\tth_1$ remain unchanged. The same process is repeated until the current filtering time $t_k$ when the last batch of observations $\ty_k$ arrives. Notice that while the non-Gaussian state distributions are allowed to evolve over time, the conjugate nature of the update ensures that the prior and posterior belong to the same distribution family.}
  \label{fig:CTF}
\end{figure}

The schematic in Fig.~\ref{fig:CTF} illustrates the sequential assimilation of new observations within \texttt{CTF}. One aspect that becomes immediately apparent is that state functions $\bvf_{\tth_k}$ are modified only during the prediction step. During the \texttt{CTF} update, the parameters of $\bvf_{\tth_k}$ remain unchanged, and the assimilated observations affect only the latent Gaussian parameters; that is, they transform $\left\{ \upmu_{k|k-1}, \bE_{k|k-1} \right\}$ to $\left\{ \upmu_{k|k}, \bE_{k|k} \right\}$.

Another important property is that the \texttt{KF} emerges as a special case of \texttt{CTF}, as formalized in the next corollary.

\vspace*{3mm}
\noindent \textbf{Corollary} (Generalization of the \texttt{KF}). \textit{If $\bvf_{\tth_k}$ and $\bvg_{\tth_k}$ are chosen to be the identity transformation, (\ref{eq:CTF_theorem-pdf})--(\ref{eq:CTF_theorem-gain}) recover the standard Kalman filter equations. In addition, if $\bvf_{\tth_k}$ and $\bvg_{\tth_k}$ are affine functions, the \texttt{CTF} update reduces to a well-known variant of the Kalman filter obtained under linear transformations of the state and observation variables \citep[e.g., see][]{snyder_2015}.}
\vspace*{3mm}

\noindent Note that the analytical tractability of the full non-Gaussian filtering density is an important distinction to other nonlinear extensions of the \texttt{KF}, such as the extended Kalman filter (\texttt{EKF}) or the unscented Kalman filter (\texttt{UKF}), which only approximate the mean and covariance of the posterior pdf \citep{simon_2006}. 

\vspace*{-3.5mm}
\subsection{Joint state-observation space formulation}

The matrix $\bH_k$ appearing in the \texttt{CTF} recursions (\ref{eq:CTF_theorem-mu})-(\ref{eq:CTF_theorem-gain}) relates state and observation variables in the latent Gaussian space. However, its construction is not straightforward unless state variables are directly observed and the functions $\bvf_{\tth_k}$, $\bvg_{\tth_k}$ are defined elementwise. For multivariate transformations, the matrix $\bH_k$ becomes unknown and can be estimated jointly with other parameters of the SSM. Nevertheless, purely statistical methods like those adopted by \citet{rangapuram_et_al_2018} and \citet{de_bezenac_et_al_2020} assume no prior knowledge of the observed dynamical system. 

In Earth system modeling, significant efforts are devoted to the formulation and implementation of complex observation operators $\bvh_k$ which are consistent with the linear-additive observation model (\ref{eq:SSM_Y-aG}). One way to take advantage of this information is through the joint state-observation space approach, where at each time step $k$, one needs to construct the extended state vector

\vspace*{1mm}
\begin{equation} \label{eq:extended_state}
   \tZ_k = \begin{bmatrix} \tX_k \\ \bvh_k(\tX_k) \end{bmatrix} \in \mathbb{R}^{N_x + N_y}.
\end{equation}
\vspace*{1mm}

\noindent It will be further assumed that the nonlinear state function $\bvf_{\tth_k}$ can be partitioned into state and observation components such that 

\vspace*{1mm}
\begin{equation} \label{eq:f}
   \tZ_k = \bvf_{\tth_k} \left( \widetilde{\tZ_k} \right) = \bvf_{\tth_k} \left( \begin{bmatrix} \widetilde{\tX_k} \\ \widetilde{\bvh_k(\tX_k)}  \end{bmatrix} \right) = \begin{bmatrix} \bvf_{\tth_k^x} \left( \widetilde{\tX_k} \right) \\ \bvf_{\tth_k^y} \left( \widetilde{\bvh_k(\tX_k)} \right) \end{bmatrix}.
\end{equation}
\vspace*{1mm}

\noindent With this in mind, $\bH_k$ in the new observation model (\ref{eq:new_SSM_Y}) becomes a simple selection matrix returning the observation part of the latent state vector $\widetilde{\tz}_k$,

\vspace*{1mm}
\begin{equation} \label{eq:H_selection}
   \bH_k \widetilde{\tz}_k = \begin{bmatrix} \bO_{N_y \times N_x} \quad \bI_{N_y} \end{bmatrix} \begin{bmatrix} \widetilde{\tx_k} \\ \widetilde{\bvh_k(\tx_k)} \end{bmatrix} = \widetilde{\bvh_k(\tx_k)},
\end{equation}
\vspace*{1mm}

\noindent where $\bO_{N_y \times N_x}$ is a matrix of zeros with dimensions $N_y \times N_x$ and $\bI_{N_y}$ is the identity matrix of size $N_y$. 

The idea of casting the DA problem in a joint state-observation space has been already leveraged in standard (Gaussian) DA methods such as the Ensemble Adjustment Kalman Filter \citep[\texttt{EAKF;}][]{anderson_2001} to circumvent the related problem of nonlinear observation operators. At first, it seems that this approach simply moves the source of nonlinearity from the observation operator to the extended state vector $\tZ_k$. Even if $\tX_k$ is a Gaussian random vector, the application of a potentially nonlinear observation operator $\bvh_k$ will make the extended vector $\tZ_k$ non-Gaussian and generate errors in the DA update. However, the \texttt{CTF} framework is specifically designed to handle this situation by choosing an appropriate transformation $\bvf_{\tth_k^y}$. If $\tX_k$ also happens to be non-Gaussian, $\bvf_{\tth_k^x}$ can be additionally made nonlinear. 

Notice that even if the marginal distributions of $\tX_k$ and $\bvh_k(\tX_k)$ are perfectly approximated by the nonlinear functions $\bvf_{\tth_k^x}$ and $\bvf_{\tth_k^y}$, the full state-observation transformation $\bvf_{\tth_k}$ may not accurately capture the joint distribution of $\tZ_k$. In principle, it is possible to use a single multivariate diffeomorphism $\bvf_{\tth_k}$ to mix $\tX_k$ and $\bvh_k(\tX_k)$, but it turns out that separating the state and observation variables comes with certain theoretical guarantees. To see this, consider the following setting: let $\tE_k^o = 0$ almost surely (a.s.), which is satisfied when $\bR_k = \bO_{N_y \times N_y}$, and take the observation function $\bvg_{\tth_k} = \bvf_{\tth_k^y}$. Under these assumptions, the observation equation (\ref{eq:new_SSM_Y}) in the joint state-observation space becomes

\vspace*{1mm}
\begin{equation} \label{eq:unbiased_obs-zero_noise}
   \tY_k \overset{^\text{a.s.}}{=\joinrel=} \bvf_{\tth_k^y} \left( \bH_k \widetilde{\tz}_k + 0 \right)   \overset{^\text{(\ref{eq:H_selection})}}{=\joinrel=} \bvf_{\tth_k^y} \left( \widetilde{\bvh_k(\tx_k)} \right) \overset{^\text{(\ref{eq:f})}}{=\joinrel=} \bvh_k(\tx_k),
\end{equation}
\vspace*{-0.5mm}

\noindent demonstrating that observations are unbiased since they are equal to the realization of the true state in observation space. Related to this finding, the next proposition shows that in the asymptotic limit $\bR_k \to \bO_{N_y \times N_y}$, the \texttt{CTF} update converges to the observations.

\vspace*{5mm}
\noindent \textbf{\hypertarget{prop:CTF_obSpace_consistency}Proposition} (Observation-space consistency of \texttt{CTF}). \textit{In the \texttt{CTF} equations (\ref{eq:CTF_theorem-pdf})--(\ref{eq:CTF_theorem-gain}), let the state $\tX_k$ be replaced with its extended version $\tZ_k$ in (\ref{eq:extended_state}), the matrix $\bH_k$ -- defined according to (\ref{eq:H_selection}), and the function $\bvf_{\tth_k}$ -- partitioned into its state and observation components, $\bvf_{\tth_k^x}$ and $\bvf_{\tth_k^y}$, as in (\ref{eq:f}). Provided that $\bvg_{\tth_k} = \bvf_{\tth_k^y}$ and $\bR_k = \bO_{N_y \times N_y}$, the \texttt{CTF} update in observation space is centered on the observed values $\ty_k$; that is, $p \left( \bH_k \tz_k | \ty_{1:k} \right) = \delta(\bH_k \tz_k -\ty_k)\ \ \forall k \in \mathbb{Z}_{>0}$.}

\section{Ensemble approximations}
\label{sec:ECTF}

Similar to the Kalman filter, a direct application of \texttt{CTF} to Earth system models is not feasible due to its inability to calculate and store the large covariance matrices appearing in (\ref{eq:CTF_theorem-mu})--(\ref{eq:CTF_theorem-gain}). However, the closed-form filtering recursions facilitate the formulation of an unbiased ensemble approximation, which will be referred to as the \textit{Ensemble Conjugate Transform Filter} (\texttt{ECTF}) hereafter. 

To be consistent with the joint state-observation space approach of \texttt{CTF}, an extended prior ensemble is first constructed at each time $k$ by applying the nonlinear observation operator $\bvh_k$ to each forecast ensemble member $\tX_k^{f,i}$, 

\vspace*{1mm}
\begin{equation} \label{eq:extended_ensMember}
   \tZ_k^{f,i} = \begin{bmatrix} \tX_k^{f,i} \\ \bvh_k \left( \tX_k^{f,i} \right) \end{bmatrix} \in \mathbb{R}^{N_x+N_y}\ \ \text{for}\ i=1:N_e.
\end{equation}
\vspace*{1mm}

\noindent It is further assumed that $\bvf_{\tth_k}$ is partitioned according to (\ref{eq:f}) and $\bvg_{\tth_k}=\bvf_{\tth_k^y}$. The latter does not only guarantee the observation-space consistency of \texttt{CTF} (see \hyperlink{prop:CTF_obSpace_consistency}{Proposition}), but also provides a practical way to construct these functions. With this in mind, the \texttt{ECTF} algorithm can be implemented in 3 general steps.

\subsection{Fitting a non-Gaussian pdf to the prior ensemble} \label{subsec:ECTF_prior_fitting}

\begin{figure}[!ht]
  \centering
  \includegraphics[width=0.7\textwidth]{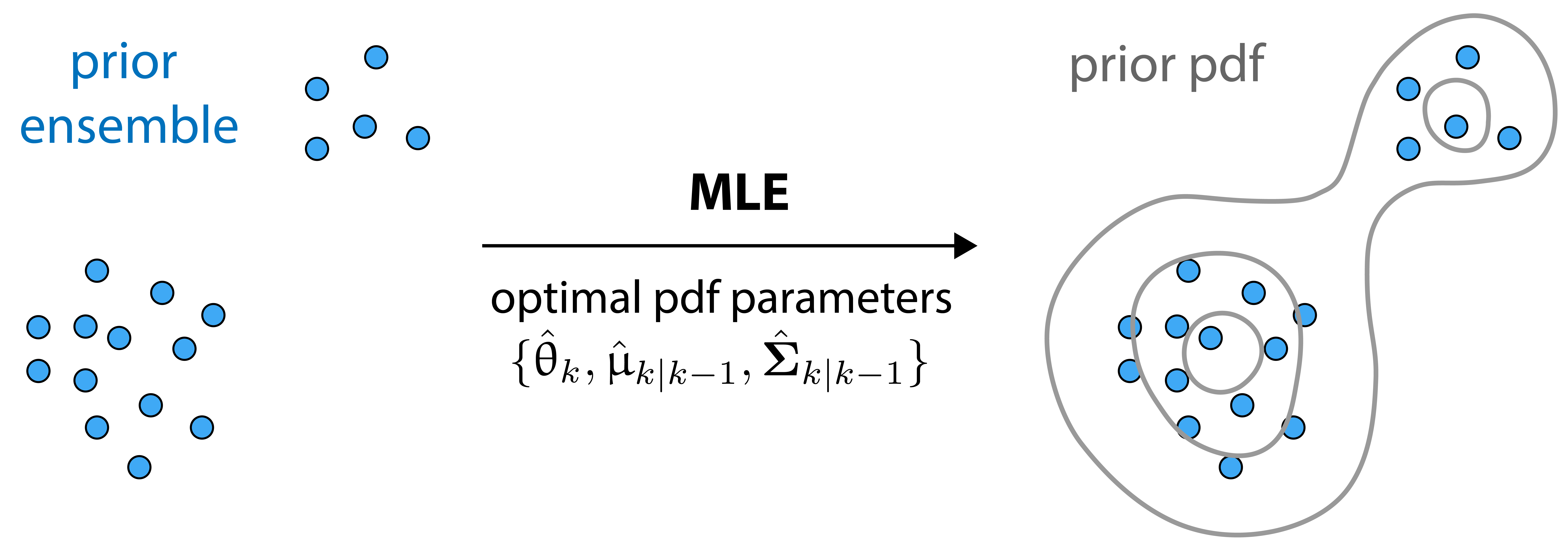}
  \caption{Using the prior ensemble $\{\tZ_k^{f,i}\}_{i=1}^{N_e}$ to fit the parameters of the non-Gaussian prior pdf $p(\tz_k | \ty_{1:k-1}) = \bvf_{\tth_k\, \sharp} \, \phi \left[\tz_k ;\upmu_{k|k-1},\bE_{k|k-1} \right]$ at time $k$.}
  \label{fig:prior_pdf_fitting}
\end{figure}

The first step is to determine which non-Gaussian pdf best describes the extended forecast ensemble $\{\tZ_k^{f,i}\}_{i=1:}^{N_e}$. According to (\ref{eq:CTF_theorem-prior_pdf}), the prior pdf $p(\tz_k | \ty_{1:k-1})$ is given by $\bvf_{\tth_k\, \sharp} \, \phi \left[\tz_k ;\upmu_{k|k-1},\bE_{k|k-1} \right]$. This means that in its most general form, the first step requires determining the optimal parameters $\{ \hat \tth_k, \hat \upmu_{k|k-1}, \hat \bE_{k|k-1} \}$ from $\{\tZ_k^{f,i}\}_{i=1}^{N_e}$. This is a classical task in Maximum Likelihood Estimation (MLE) and is illustrated in Fig.~\ref{fig:prior_pdf_fitting} for a bimodal prior ensemble in $\mathbb{R}^2$.

Clearly, the choice of $\bvf_{\tth_k}$ plays a crucial role for specifying the non-Gaussian characteristics of the prior pdf. Since the only restriction is that this function is a diffeomorphism, there is considerable freedom in how it is designed. In the simplest case where the system dynamics are known and weakly non-Gaussian, the DA practitioner may choose to use elementwise nonlinearities with \textit{fixed} parameters $\tth_k$ (or no extra parameters at all). In this situation, one only needs to calculate the optimal parameters $\{ \hat \upmu_{k|k-1}, \hat \bE_{k|k-1} \}$ associated with the reference Gaussian. Similar to the \texttt{EnKF}, this can be done analytically by transforming $\{\tZ_k^{f,i}\}_{i=1}^{N_e}$ with $\bvf_{\tth_k}^{-1}$ and calculating the sample mean and covariance of the resulting sample $\{\tilde{\tZ}_k^{f,i}\}_{i=1}^{N_e}$\footnote{In \texttt{EnKF}, the prior pdf is Gaussian and $\bvf_{\tth_k}$ is simply the identity transformation. In this situation, there is no need to apply $\bvf_{\tth_k}^{-1}$ before calculating the sample mean and covariance.}. If the system dynamics are highly non-Gaussian and only partially known, a more appropriate choice would be to stack (compose) multiple nonlinearities with \textit{learnable} parameters $\tth_k$. In addition to its functional representations, $\bvf_{\tth_k}$ can be either estimated \textit{online} (at each analysis step $k$) or \textit{offline} (e.g., based on historical ensemble simulations). For example, complex neural network parameterizations of $\bvf_{\tth_k}$ are expected to incur significant computational costs, which makes them more suitable for offline estimation (learning).

\subsection{Updating the latent Gaussian parameters} \label{subsec:ECTF_latent_update}

Having determined the prior pdf $p(\tz_k | \ty_{1:k-1})$ and the corresponding functions $\bvf_{\tth_k}$ and $\bvg_{\tth_k}$, the next step is to map variables into the latent Gaussian space and perform an update based on the current observations $\ty_k$. Since (\ref{eq:CTF_theorem-mu})--(\ref{eq:CTF_theorem-gain}) represent a \texttt{KF}-like recursion, any available \texttt{EnKF} solver can be used to obtain unbiased estimates of $\{\upmu_{k|k},\bE_{k|k}\}$ from the latent-space analysis ensemble $\{\tilde{\tZ}_k^{a,i} \}_{i=1}^{N_e}$. For instance, applying the stochastic \texttt{EnKF} formula \citep{vanLeeuwen_2020} gives

\vspace*{-3mm}
\begin{align}
    & \tilde{\tZ}_k^{a,i} = \tilde{\tZ}_k^{f,i} + \hat{\bK}_k \left[ \tilde \ty_k - \bH_k \tilde{\tZ}_k^{f,i} - \tE_k^{o,i} \right]\ \text{for}\ \ i=1:N_e \label{eq:EnKF_stochastic-ens},\\
    & \hat{\bK}_k = \hat{\bE}_{k|k-1} \bH_k^\top \left( \bH_k \hat{\bE}_{k|k-1} \bH_k^\top + \hat \bR_k  \right)^{-1} \label{eq:EnKF_stochastic-gain},
\end{align}
\vspace*{-3mm}

\noindent where $\tilde{\tZ}_k^{f,i} = \bvf_{\hat \tth_k}^{-1}({\tZ}_k^{f,i})$, $\tilde \ty_k = \bvf_{\hat \tth_k^y}^{-1} (\ty_k)$ and $\{\tE_k^{o,i}\}_{i=1}^{N_e}$ are $N_e$ realizations of the observation noise $\tE_k^{o} \sim \mathcal{N}(0,\hat \bR_k)$. Similar to other \texttt{EnKF}s, the Kalman gain terms involving $\hat{\bE}_{k|k-1}$ can be computed more efficiently \citep[see Section 2a of][]{houtekamer_zhang_2016} from 

\vspace*{-1mm}
\begin{align}
    & \hat{\bE}_{k|k-1} \bH_k^\top = \textbf{Cov}^{(N_e)} \left[ \{\tilde{\tZ}_k^{f}\}, \{\bH_k \tilde{\tZ}_k^{f}\} \right] \label{eq:EnKF_stochastic-PbHt},\\
    & \bH_k \hat{\bE}_{k|k-1} \bH_k^\top = \textbf{Cov}^{(N_e)} \left[ \{\bH_k \tilde{\tZ}_k^{f}\}, \{\bH_k \tilde{\tZ}_k^{f}\} \right] \label{eq:EnKF_stochastic-HPbHt},
\end{align}
\vspace*{1mm}

\noindent where $\textbf{Cov}^{(N)}[\cdot,\cdot]$ is the $N$-sample covariance operator defined as

\vspace*{1mm}
\begin{equation}
    \textbf{Cov}^{(N)} \left[ \{\tW_1\}, \{\tW_2\} \right] \coloneqq \frac{1}{N-1} \sum_{i=1}^{N} \left( \tW_1^i- \frac{1}{N}\sum_{j=1}^{N} \tW_1^j \right) \left( \tW_2^i- \frac{1}{N}\sum_{j=1}^{N} \tW_2^j \right)^\top.
\end{equation}

If $\hat \bR_k$ is fixed \textit{a priori}, one can proceed with calculating $\hat{\bK}_k$, simulating $\{\tE_k^{o,i}\}_{i=1}^{N_e}$ and completing the stochastic EnKF update in (\ref{eq:EnKF_stochastic-ens}). However, even if there was perfect knowledge of the observation error statistics in physical space, it is not straightforward to relate them to the covariance matrix of the observation noise in the latent Gaussian space. 

A more objective approach to estimate $\hat \bR_k$ is from the perturbed observation ensemble $\{\tY_k^{f,i}\}_{i=1}^{N_e}$, which can be generated by evaluating $p(\ty_k|\tz_k)$ at each ensemble member $\tZ_k^{f,i}$. If $p(\ty_k|\tz_k)$ is not explicitly available, it is still possible to use an oracle/simulator for the observation model \citep{taghvaei_hosseini_2022}. The important point in both cases is that $\{\tY_k^{f,i}\}_{i=1}^{N_e}$ is a sample from $p(\ty_k|\ty_{1:k-1})$. Under the new SSM (\ref{eq:new_SSM}), this pdf is given by

\vspace*{1mm}
\begin{equation}
    p(\ty_k|\ty_{1:k-1}) = \bvg_{\tth_k\, \sharp} \, \phi \left[\ty_k; \bH_k \upmu_{k|k-1},\bH_k \bE_{k|k-1} \bH_k^\top + \bR_k \right].
\end{equation}
\vspace*{-1mm}

\noindent The form of $p(\ty_k|\ty_{1:k-1})$ follows from the non-Gaussian pdf product \hyperlink{lemma:prod_nonGauss_pdfs}{Lemma} applied to the prior $p(\tz_k|\ty_{1:k-1})$ and likelihood $p(\ty_k|\tz_k)$ in (\ref{eq:CTF_theorem-prior_pdf}) and (\ref{eq:cond_pdfs-new_SSM-ob}), respectively. Since in our setting $\bvg_{\tth_k}=\bvf_{\tth_k^y}$, using the inverse function $\bvf_{\tth_k^y}^{-1}$ yields $\{\tilde \tY_k^{f,i}\}_{i=1}^{N_e}$, the perturbed observation ensemble in the latent Gaussian space. After this is done, the inverse term in the Kalman gain (\ref{eq:EnKF_stochastic-gain}) can be estimated from

\vspace*{1mm}
\begin{equation}
    \bH_k \hat{\bE}_{k|k-1} \bH_k^\top + \hat \bR_k  = \textbf{Cov}^{(N_e)} \left[ \{\tilde{\tY}_k^{f}\}, \{\tilde{\tY}_k^{f}\} \right].
\end{equation}
\vspace*{-1mm}

\noindent Generating $\{\tE_k^{o,i}\}_{i=1:N_e}$ then pertains to sampling from a 0-mean Gaussian with covariance $\hat \bR_k = \textbf{Cov}^{(N_e)} \left[ \{\tilde{\tY}_k^{f}\}, \{\tilde{\tY}_k^{f}\} \right] - \bH_k \hat{\bE}_{k|k-1} \bH_k^\top$, where the second term is given by (\ref{eq:EnKF_stochastic-HPbHt}). An alternative solution would be to realize that according to the new observation model (\ref{eq:new_SSM_Y}),

\begin{equation}
    \bH_k \tilde{\tZ}_k^{f,i} + \tE_k^{o,i} = \tilde{\tY}_k^{f,i}.
\end{equation}
\vspace*{-1mm}

\noindent This implies that once $\{\tilde{\tY}_k^{f,i}\}_{i=1}^{N_e}$ is obtained, completing the stochastic \texttt{EnKF} analysis does not require the explicit estimation of $\hat \bR_k$ or the generation of $\{\tE_k^{o,i}\}_{i=1}^{N_e}$. Note that the numerical implementation of \texttt{ECTF} in Section \ref{sec:exps} follows this alternative strategy.

\subsection{Transforming back to physical space} \label{subsec:ECTF_backtransforming}

The final step in \texttt{ECTF} is informed by the form of the posterior pdf, $p(\tz_k | \ty_{1:k}) = \bvf_{\tth_k\, \sharp} \, \phi \left[\tz_k ;\upmu_{k|k},\bE_{k|k} \right]$. Notice that all calculations in the previous subsection were designed to generate a sample $\{\tilde{\tZ}_k^{a,i} \}_{i=1}^{N_e}$ from the base Gaussian density in this posterior. Therefore, to get a sample from the full posterior, the state function $\bvf_{\hat \tth_k}$ determined in Section \ref{subsec:ECTF_prior_fitting} is evaluated at each member of $\{\tilde{\tZ}_k^{a,i} \}_{i=1}^{N_e}$; that is,

\vspace*{1mm}
\begin{equation}
    \tZ_k^{a,i} = \bvf_{\hat \tth_k} (\tilde{\tZ}_k^{a,i}) \quad \text{for}\ i=1:N_e.
\end{equation}
\vspace*{-1mm}

\noindent As the statistical experiments from Section \ref{sec:exps} will demonstrate, the resulting sample $\{\tZ_k^{a,i} \}_{i=1}^{N_e}$ provides an unbiased approximation of the true \texttt{CTF} update.

\subsection{Relation to other nonlinear ensemble filters} \label{subsec:ECTF_relations}

This section concludes with brief comments on the connection between \texttt{ECTF} and two other nonlinear ensemble filters. The first one is the Gaussian anamorphosis \texttt{EnKF} \citep[\texttt{GA-EnKF;}][]{bertino_et_al_2003,simon_bertino_2009}, which also leverages separate transformations of state and observation variables together with a \texttt{KF}-like update in the latent Gaussian space. However, one distinction is that most \texttt{GA-EnKF} algorithms are based on a different update in the latent Gaussian space, whereby the observation operator $\bH_k$ is composed with the nonlinear functions $\bvf_{\tth_k}$ and $\bvg_{\tth_k}$ \citep[e.g., see eq.~4 of][]{simon_bertino_2009}. Unfortunately, the theoretical properties of this formulation have not been investigated in previous studies. Another common drawback of \texttt{GA-EnKF} methods is that they use elementwise transformations which limit their ability to represent more complex posterior distributions. However, an even more critical difference is the ambiguity in constructing the anamorphosis functions themselves. Typically, standard Gaussian references are chosen, but \citet{amezcua_vanLeeuwen_2014} apply moment-matching Gaussians. By contrast, \texttt{ECTF} offers a principled way to estimate the latent Gaussian parameters of the prior and likelihood directly from the augmented forecast ensemble $\{\tZ_k^{f,i}\}_{i=1:}^{N_e}$ (\textit{cf}.~Section \ref{subsec:ECTF_latent_update}).

The change-of-variables theorem used to derive \texttt{ECTF} is closely related to the broader area of measure transport. Tools from measure transport have already been used to formulate non-Gaussian ensemble DA methods, such as the Stochastic Map Filter (\texttt{SMF}) of \citet{spantini_et_al_2022}. Similar to \texttt{ECTF}, \texttt{SMF} has the capability to work with multivariate nonlinear functions, which are parameterized and estimated from the joint state-observation ensemble. The reference distribution in \texttt{SMF} is fixed to a standard Gaussian, making this method algorithmically similar to the \texttt{GA-EnKF} approach. Unlike \texttt{GA-EnKFs}, however, the \texttt{SMF} analysis is guaranteed to be Bayesian consistent and gives increasingly closer approximations to the true posterior with larger ensemble sizes. It should be also noted that while \texttt{SMF}s do not make any assumptions on the prior and likelihood, they are restricted to triangular (Knothe-Rosenblatt) maps. On the other hand, \texttt{ECTF} imposes parametric assumptions on the prior and likelihood (see eqs.~\ref{eq:CTF_theorem-prior_pdf} and \ref{eq:cond_pdfs-new_SSM-ob}), but can be implemented with a wider class of nonlinear transformations. The latter might include complex neural network architectures, in which case \texttt{ECTF} is in a position to take advantage of efficient AI-based optimization libraries.

\section{Statistical experiments}
\label{sec:exps}

In this section, idealized statistical experiments are performed with two specific objectives in mind: (i) to validate the Bayesian consistency of \texttt{ECTF} and (ii) to illustrate the advantages of its multivariate non-Gaussian update. To this end, \texttt{ECTF}'s performance is compared against the standard \texttt{EnKF}, which makes fully Gaussian assumptions, and the recently introduced Quantile-Conserving Ensemble Filter (\texttt{QCEF}) framework of \citet{anderson_2022}. The implementation of \texttt{EnKF} is analogous to that of \texttt{ECTF} (\textit{cf}.~Section \ref{sec:ECTF}), but the state function $\bvf_{\tth_k}$ is set to the identity transformation. On the other hand, the \texttt{QCEF} framework is a two-step approach which first uses a quantile-matching procedure to update the prior ensemble in observation space \citep[eq.~2 in][]{anderson_2022} and then regresses the corresponding analysis increments to the remaining state variables. While the formulation of the second step is quite flexible, the experiments in this study use a standard Linear Regression \citep[LR; see eq.~5 in][]{anderson_2003}, which implicitly makes Gaussian assumptions on the posterior distribution of the unobserved state variables \citep{grooms_2022}. To stress the linear nature of the second step, the resulting method is abbreviated as \texttt{QCEF-LR} and its main intention is to facilitate a smoother transition between the fully Gaussian and non-Gaussian updates of \texttt{EnKF} and \texttt{ECTF}.

\subsection{Bayesian setting} \label{subsec:exps_bayesian_setting}

The idealized statistical experiments are based on a two-dimensional Bayesian problem in which the prior state at time $k$, $\tZ_k = \left[\tZ_{1,k}\ \tZ_{2,k} \right]^\top$, is distributed according to

\begin{equation} \label{eq:exps_prior}
    p(\tz_k|\ty_{1:k-1}) = \frac{1}{2\pi\ (det\ \bE_{k|k-1})^{1/2}} exp\left[ -\frac{1}{2} || \bvf_k^{-1}(\tz_{1,k},\tz_{2,k})-\upmu_{k|k-1}||_{\bE_{k|k-1}}^2 \right] \frac{1}{\tz_{1,k} \tz_{2,k} (1-\tz_{2,k})},
\end{equation}

\noindent where $||\tx||^2_\bE \coloneqq \tx^\top \bE^{-1} \tx$ and $\bvf_k^{-1}: (0,\infty) \times (0,1) \to \mathbb{R}^2$ is given by

\vspace*{1mm}
\begin{equation}
    \bvf_k^{-1}(\tz_{1,k},\tz_{2,k}) = \begin{bmatrix} ln(\tz_{1,k}) \\ ln \left( \frac{\tz_{2,k}}{1-\tz_{2,k}} \right) \end{bmatrix} =: \tilde \tz_k.
\end{equation}
\vspace*{1mm}

\noindent This prior has the form of (\ref{eq:CTF_theorem-prior_pdf}) -- a Gaussian pdf pushed forward by a \textit{non-parametric elementwise} nonlinearity $\bvf_k$ whose first and second component are the exponential and standard logistic functions\footnote{The standard logistic function is defined as $x \mapsto \frac{1}{1+exp(-x)}$ for $x \in \mathbb{R}$.}. The terms in the fraction to the right of the square bracket come from the volume correction associated with $\bvf_k$. It is worth noting that this prior is similar to the hybrid Gaussian-lognormal pdf of \citet{fletcher_zupanski_2006b}, but the Gaussian marginal is replaced with a logit-normal pdf which is bounded between $0$ and $1$.

In all data assimilation experiments, observations $y_k$ are made only of the first state variable $\tz_{1,k}$. In this case, $\bvf_k$ is already partitioned into state and observation components (see eq.~\ref{eq:f}) such that the observation operator simplifies to $\bH_k = [1\ 0]$ and automatically satisfies the linearity requirement of \texttt{CTF}. To enforce analysis consistency in observation space (\textit{cf.}~\hyperlink{prop:CTF_obSpace_consistency}{Proposition}), one needs to additionally set $g_k(\cdot) = exp(\cdot)$, which induces the likelihood

\vspace*{1mm}
\begin{equation}
    p(y_k|\tz_k) = \frac{1}{\sqrt{2\pi r}}\ exp\left \{ -\frac{\left[ ln(y_k) - \bH_k \tilde \tz_k \right]^{2}}{2 r} \right \} \frac{1}{y_k}.
\end{equation}
\vspace*{1mm}

The Bayesian problem described so far is especially suitable for exploring differences in \texttt{EnKF}, \texttt{QCEF-LR} and \texttt{ECTF}. Its low dimensional structure allows for a direct evaluation of Bayes' theorem. At the same time, it is sufficiently complex to study the multivariate aspects of different ensemble filters. The parametric forms of the prior and likelihood are consistent with \texttt{CTF} theory, and the resulting posteriors serve as a clear benchmark to validate the Bayesian consistency of \texttt{ECTF}. The choice of elementwise transformations in $\bvf_k$ is also deliberate: these simpler nonlinearities give access to the marginal cumulative distribution functions (cdfs) of the prior and posterior, which are needed for \texttt{QCEF-LR}'s implementation \citep[see eq.~2 of][]{anderson_2022}. 

Besides these experimental considerations, the simulated non-Gaussianity in our problem reflects common challenges in Earth system models, such as the presence of physical bounds. The types of prior and posterior pdfs (e.g., see Fig.~\ref{fig:example}) have also appeared in previous studies. Examples include Fig.~5 of \citet{posselt_vukicevic_2010} and Figs.~10a-d of \citet{poterjoy_2022}, where similar distributions arise either due to nonlinear relationships between microphysical parameters or misalignment errors in tropical cyclones.  

\subsection{Multiple DA trials} \label{subsec:multiple_trials}

The ensemble filtering performance is evaluated as a function of \textit{observation accuracy} and \textit{state correlations}. The observation accuracy is informed by the parameters of $p(y_k|\tz_k)$. The choice of a lognormal likelihood is convenient since there exists an analytical expression for the variance of observations given the state,

\vspace*{1mm}
\begin{equation}
    \text{Var}[\tY_k|\tZ_k] = [exp(r)-1]\ exp(2 \bH_k \tilde \tz_k + r).
\end{equation}
\vspace*{1mm}

\noindent This means that the magnitude of the observation errors can be controlled by modifying $r$, the variance in the latent Gaussian space. 

Similarly, the special structure of the prior pdf (\ref{eq:exps_prior}) enables adjustments to the state correlations via the covariance matrix $\bE_{k|k-1}$, which can be written as 

\vspace*{1mm}
\begin{equation}
    \bE_{k|k-1} = \begin{bmatrix} \sigma_1&  \rho \sqrt{\sigma_1 \sigma_2} \\ \rho \sqrt{\sigma_2 \sigma_1}& \sigma_2 \end{bmatrix}
\end{equation}
\vspace*{1mm}

\noindent for $\sigma_1,\sigma_2 > 0$ and $\rho \in [-1,1]$. Increasing the absolute value of $\rho$ strengthens the correlations in the latent Gaussian space, which in turn results in a more strongly and nonlinearly correlated prior ensemble. In the extreme case when $\rho = 0$, the latent Gaussian variables are uncorrelated and, by the special properties of Gaussian distributions, independent. A simple probability argument can be further applied to demonstrate that elementwise transformations of independent random variables preserve independence. Hence, setting $\rho = 0$ will produce a prior ensemble with two independent state components.  

For each parameter pair $\{\rho,r\}$, 1000 independent DA trials are performed to achieve a robust comparison between the three ensemble filters. The prior ensemble $\{\tZ_k^{f,i}\}_{i=1:}^{N_e}$ consists of $N_e=10^6$ members and is generated by sampling from  $p(\tz_k|\ty_{1:k-1})$ with $\upmu_{k|k-1}$ and $\sigma_{\{1,2\}}$ drawn from two uniform distributions with ranges $[-1,1]$ and $[0.05,2]$, respectively. The prior pdf is also used to generate the true state $\tz_k$, which serves as input to the lognormal observation model

\vspace*{0.5mm}
\begin{equation}
    y_k = exp \left( \bH_k \tilde \tz_k + \mathcal{N}(0,r) \right) = \tz_{1,k} \mathcal{LN}(0,r),
\end{equation}
\vspace*{0.5mm}

\noindent where, with a slight abuse of notation, $\mathcal{N}(0,r)$ and $\mathcal{LN}(0,r)$ denote Gaussian and lognormal random variables. 

\subsection{Skill metrics} \label{subsec:exps_skill_metrics}

The skill of different ensemble filters is based on the true posterior distribution $\pi_{\tZ_k|\tY_{1:k}}$, which is computed numerically by discretizing the two state components. In particular, $z_{1,k}$ and $z_{2,k}$ take values in $[10^{-15},500]$ and $[10^{-15},1-10^{-15}]$, with the total number of grid points set to $N_{z_1} = 250,000$ and $N_{z_2} = 100$. Similar numerical evaluations of the posterior pdf were also carried out by \citet{suselj_et_al_2020} but in the context of estimating parameters of physical processes in numerical models.

Once $\pi_{\tZ_k|\tY_{1:k}}$ is obtained, the analysis ensemble associated with each filter is converted to a histogram $\hat \pi_{\tZ_k|\tY_{1:k}}$ whose bins are centered over the discretized grid points and allow for a direct comparison with the true Bayesian posterior. The statistical distance between the two distributions is measured via the Jensen-Shannon divergence,

\vspace*{1mm}
\begin{equation} \label{eq:JS_div}
    \text{JS}(\hat \pi_{\tZ_k|\tY_{1:k}}||\pi_{\tZ_k|\tY_{1:k}}) \coloneqq \frac{1}{2}[\text{KL}(\hat \pi_{\tZ_k|\tY_{1:k}}||\pi_{\text{avg}}) + \text{KL}(\pi_{\tZ_k|\tY_{1:k}}||\pi_{\text{avg}})].
\end{equation}
\vspace*{1mm}

\noindent In the above expression, $\text{KL}(\cdot,\cdot)$ refers to the Kullback-Leibler divergence, which in our discretized setting is defined as

\vspace*{1mm}
\begin{equation} \label{eq:KL_div}
    \text{KL}(p||q) \coloneqq \sum_{i=1}^{N_{z_1}} \sum_{j=1}^{N_{z_2}} p_{i,j}\ ln\left( \frac{p_{i,j}}{q_{i,j}} \right)
\end{equation}
\vspace*{1mm}

\noindent for any two distributions $p$ and $q$ whose subscripts indicate different locations in the discretized state space. Like KL, the JS divergence takes non-negative values with 0 corresponding to a perfect distributional match, but has the additional benefit of being symmetric; i.e., $\text{JS}(\hat \pi_{\tZ_k|\tY_{1:k}}||\pi_{\tZ_k|\tY_{1:k}}) = \text{JS}(\pi_{\tZ_k|\tY_{1:k}}||\hat \pi_{\tZ_k|\tY_{1:k}})$.

The availability of a true Bayesian posterior also makes it possible to derive more traditional moment-based skill metrics, such as the mean error (ME) in the analysis mean $\upmu^a \coloneqq \upmu^a_{k|k}$ and analysis standard deviation $\upsigma^a \coloneqq \text{diag}(\bE_{k|k})$\footnote{The function $\text{diag}(\cdot)$ extracts the diagonal of a matrix.}. Using the hat notation to denote ensemble-based estimators and subscripts to indicate dimension, these two skill metrics are defined as

\vspace*{1mm}
\begin{align}
    & \text{ME}(\upmu^a) \coloneqq \frac{1}{2} \sum_{i=1}^{2} (\hat \upmu^a_i - \upmu^a_i),\\
    & \text{ME}(\upsigma^a) \coloneqq \frac{1}{2} \sum_{i=1}^{2} (\hat \upsigma^a_i - \upsigma^a_i).
\end{align}
\vspace*{1mm}

Given the bounded character of our Bayesian problem, it is also useful to quantify the percentage of analysis members $\{\tZ_k^{a,i}\}_{i=1}^{N_e}$ violating the physical bounds. Unlike the previous three skill metrics, this calculation does not require access to the true Bayesian posterior.  

\subsection{Results} \label{subsec:exps_results}

\subsubsection{Dependence of filtering skill on the observation errors and state correlations.} 

\begin{figure}[!ht]
  \centering
  \includegraphics[width=0.98\textwidth]{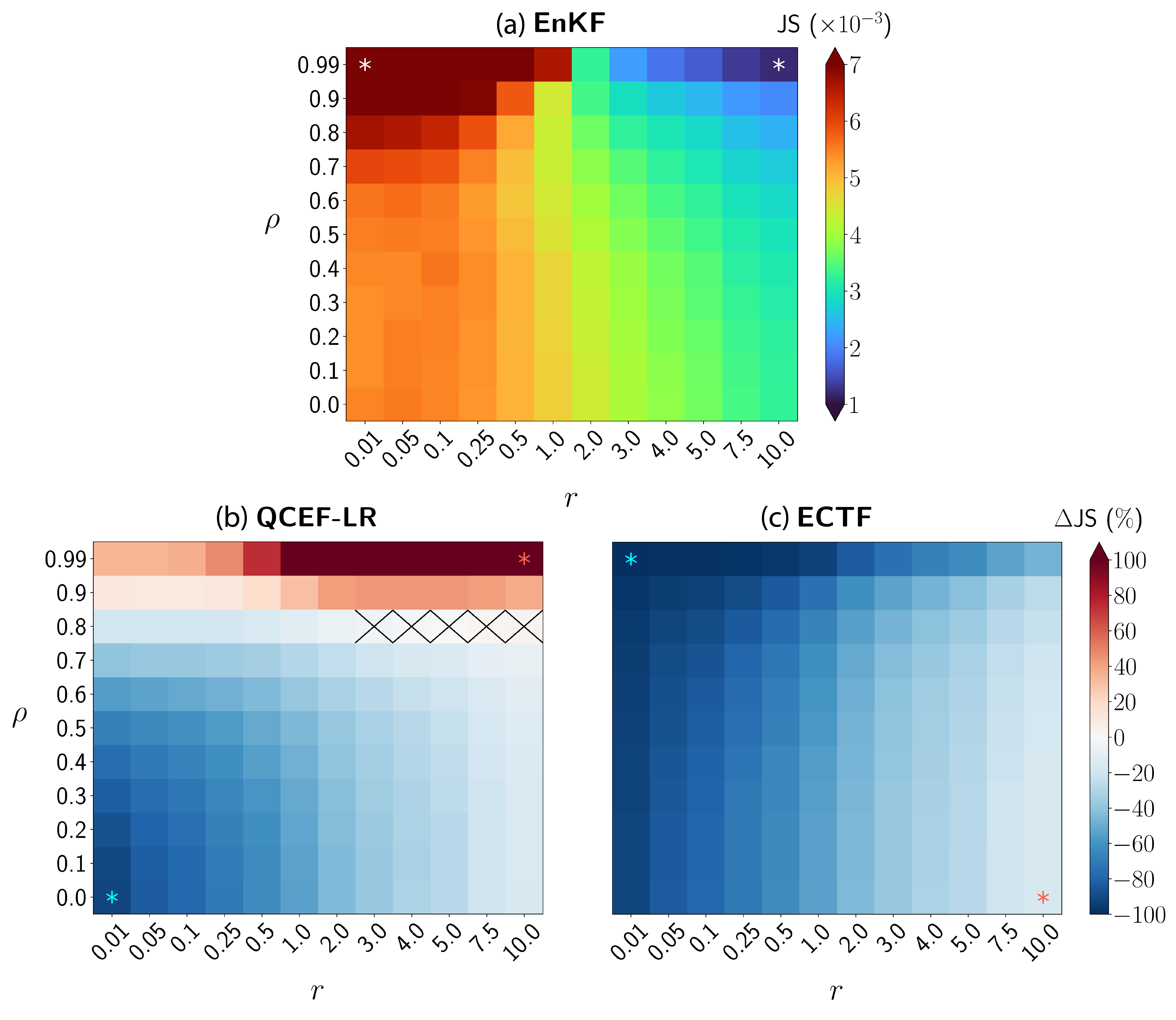}
  \caption{Ensemble filtering skill in terms of the Jensen-Shannon (JS) divergence. Panel (a) shows the scaled JS values for \texttt{EnKF}, whereas (b) and (c) -- percentage changes in this metric for \texttt{QCEF-LR} and \texttt{ECTF}. For each parameter pair $\{ \rho,r \}$, the ensemble filtering performance is averaged over 1000 independent DA trials. The black crosses in (b) indicate parameter values where the mean JS differences with \texttt{EnKF} are \textit{not} statistically significant at the 95\% confidence level according to a two-sided t-test. The lack of black crosses in (c) indicates that all ECTF-EnKF differences are statistically significant.}
  \label{fig:rho_r}
\end{figure}

The main findings from the multiple-trial DA experiments are summarized in Fig.~\ref{fig:rho_r} and presented with respect to the JS divergence. Examining the \texttt{EnKF} skill in Fig.~\ref{fig:rho_r}a, a clear degradation (higher JS values) is evident with increasing observation accuracy (smaller $r$ values). This effect is especially pronounced when the prior state is highly correlated, with \texttt{EnKF} reaching its best and worst performance for $\rho=0.99$ (top row). The fact that \texttt{EnKF} exhibits the most significant deviations from the true posterior when $r=0.01$ and $\rho=0.99$ is expected: more accurate observations cause larger analysis increments and increasing the value of $\rho$ results in a stronger nonlinear relationship between the two state variables. In both cases, it becomes increasingly more important to use a nonlinear/non-Gaussian filtering update. 

Another interesting result emerges if the \texttt{EnKF} skill is tracked for fixed values of $r$. When $r \ge 2$, JS divergences decrease with increasing $\rho$, whereas if $r < 2$, the opposite effect takes place. In the first regime, \texttt{EnKF} leads to relatively small analysis increments as a result of the large observation errors. At the same time, the higher prior correlations lead to tighter pdfs such that the cumulative differences between the ensemble-based and true posteriors (which define the JS and KL divergences in eqs.~\ref{eq:JS_div} and \ref{eq:KL_div}) become smaller. When combined, these two effects explain the small JS values in the upper-right corner of Fig.~\ref{fig:rho_r}a. When $r < 2$, the differences between the prior and posterior grow due to the assimilation of more accurate observations. The consequence of having tighter pdfs when $\rho$ increases has the opposite impact here, as it becomes less likely to have an overlap between the biased \texttt{EnKF} posterior and the true Bayesian solution.

Shifting our attention to Fig.~\ref{fig:rho_r}b, it is clear that \texttt{QCEF-LR} shows improvements over \texttt{EnKF} for most parameter choices, with the greatest benefits evident in the low $\rho$/low $r$ regime. The advantages of \texttt{QCEF-LR} in the case of weak state correlations can be explained by considering the limit $\rho \rightarrow 0$. As discussed in Section \ref{subsec:multiple_trials}, the two state variables are independent and uncorrelated in this case, which implies no transfer of observational information from observed to unobserved variables. Given that \texttt{QCEF-LR} provides an exact analysis for the $z_{1,k}$ variable by construction, the end result is a Bayesian consistent update in this joint 2-dimensional state space.

Nevertheless, \texttt{QCEF-LR} degradations are found when the prior correlations are strong ($\rho > 0.8$), especially in the case of error-prone observations (upper right corner of Fig.~\ref{fig:rho_r}b) when prior-to-posterior differences should be small. While both \texttt{EnKF} and \texttt{QCEF-LR} rely on a linear regression to update $z_{2,k}$ (the unobserved state variable), \texttt{QCEF-LR} does not explicitly account for observation errors in the second update step. Indeed, using eq.~(5) from \citet{anderson_2003} shows that it is possible to obtain non-negligible analysis increments in $z_{2,k}$ when the prior variance in $z_{1,k}$ is relatively small (which will happen when the prior ensemble is close to the 0 bound).

Finally, Fig.~\ref{fig:rho_r}c shows that \texttt{ECTF} leads to systematically better results compared to \texttt{EnKF} over the entire parameter space. The most significant improvements occur for high $\rho$ and low $r$ values (upper left corner of Fig.~\ref{fig:rho_r}c), which is also the regime where \texttt{EnKF} performs the worst. Unlike \texttt{QCEF-LR}, the best \texttt{ECTF} results are achieved when the prior ensemble is highly correlated since \texttt{ECTF} can effectively handle nonlinear relationships between state variables. 

\subsubsection{Further exploration of the $\{\rho=0.99,r=0.01\}$ regime.}

Additional experiments are performed for $\rho=0.99$ and $r=0.01$ as these parameter choices reflect the regime where \texttt{ECTF} exhibits the largest advantages over \texttt{EnKF}. Specifically, multiple DA trials are conducted with fixed observation values $y_k \in [0.5,20]$ and the results are then examined as a function of the innovations 

\vspace*{1mm}
\begin{equation} \label{eq:innovation}
    d_k = y_k - \frac{1}{N_e}\sum_{i=1}^{N_e} \bH_k \tZ_k^{f,i}.
\end{equation}
\vspace*{1mm}

\begin{figure}[!ht]
  \centering
  \includegraphics[width=0.8\textwidth]{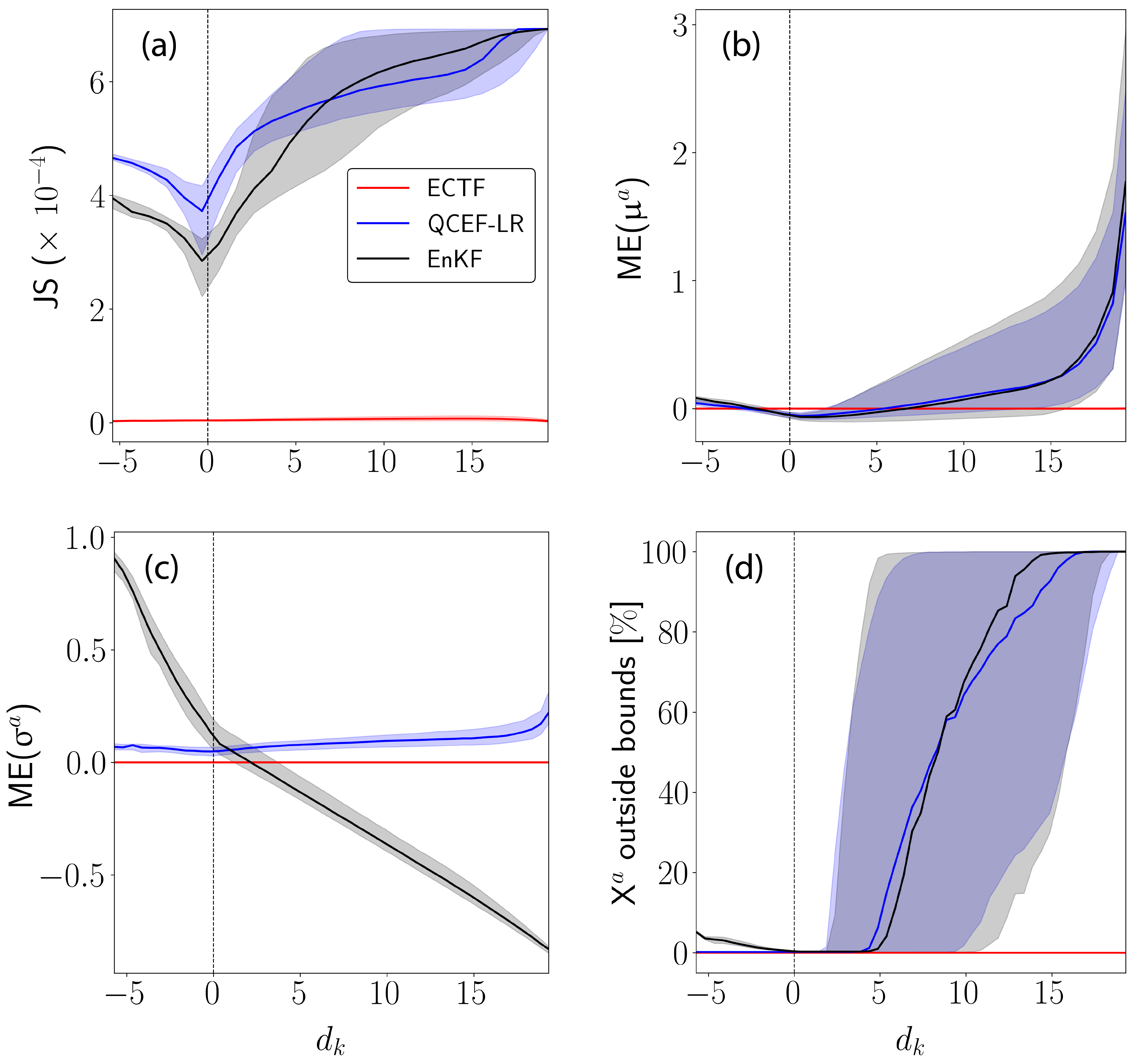}
  \caption{A closer examination of the $\{\rho=0.99,r=0.01\}$ parameter regime. Panels (a)--(d) display the median performance of \texttt{EnKF} (black), \texttt{QCEF-LR} (blue) and \texttt{ECTF} (red) as a function of the innovation values $d_k$ in (\ref{eq:innovation}) and with respect to all 4 skill metrics described in Section \ref{subsec:exps_skill_metrics}. The shading around each curve indicates the interquartile range (IQR) calculated from the multiple trial DA experiments (hard to distinguish for \texttt{ECTF} due to the small variability in its skill).}
  \label{fig:skill_metrics}
\end{figure}

The JS divergences in Fig.~\ref{fig:skill_metrics}a confirm the results from Fig.~\ref{fig:rho_r}, but additionally reveal that the largest error contributions in \texttt{EnKF} and \texttt{QCEF-LR} occur when $d_k$ is away from $0$. This result is largely reproduced by the remaining panels displaying $\text{ME}(\upmu^a)$, $\text{ME}(\upsigma^a)$ and the percentage of analysis members outside the physical bounds. However, it is interesting to note that \texttt{EnKF} and \texttt{QCEF-LR} do not achieve unbiased posterior means for $d_k = 0$, but instead when $d_k \approx \{-2,5\}$ (Fig.~\ref{fig:skill_metrics}b). The behavior of $\text{ME}(\upsigma^a)$ in Fig.~\ref{fig:skill_metrics}c is also noteworthy: \texttt{QCEF-LR} consistently overestimates the posterior spread, whereas \texttt{EnKF} is overdispervise (underdispersive) for negative (positive) $d_k$ values. Perhaps surprisingly, the last panel (Fig.~\ref{fig:skill_metrics}d) does not indicate large differences between \texttt{EnKF} and \texttt{QCEF-LR} in terms of the percentage of analysis members outside the physical bounds: \texttt{EnKF} clearly performs worse when $d_k < 0$ values, but both filters show a rapid increase for $d_k > 4$. Finally, it is worth noting that all skill metrics in Fig.~\ref{fig:skill_metrics} validate the Bayesian consistency of \texttt{ECTF}.

\subsubsection{Example illustration of the filtering differences.}

\begin{figure}[!ht]
  \centering
  \includegraphics[width=0.98\textwidth]{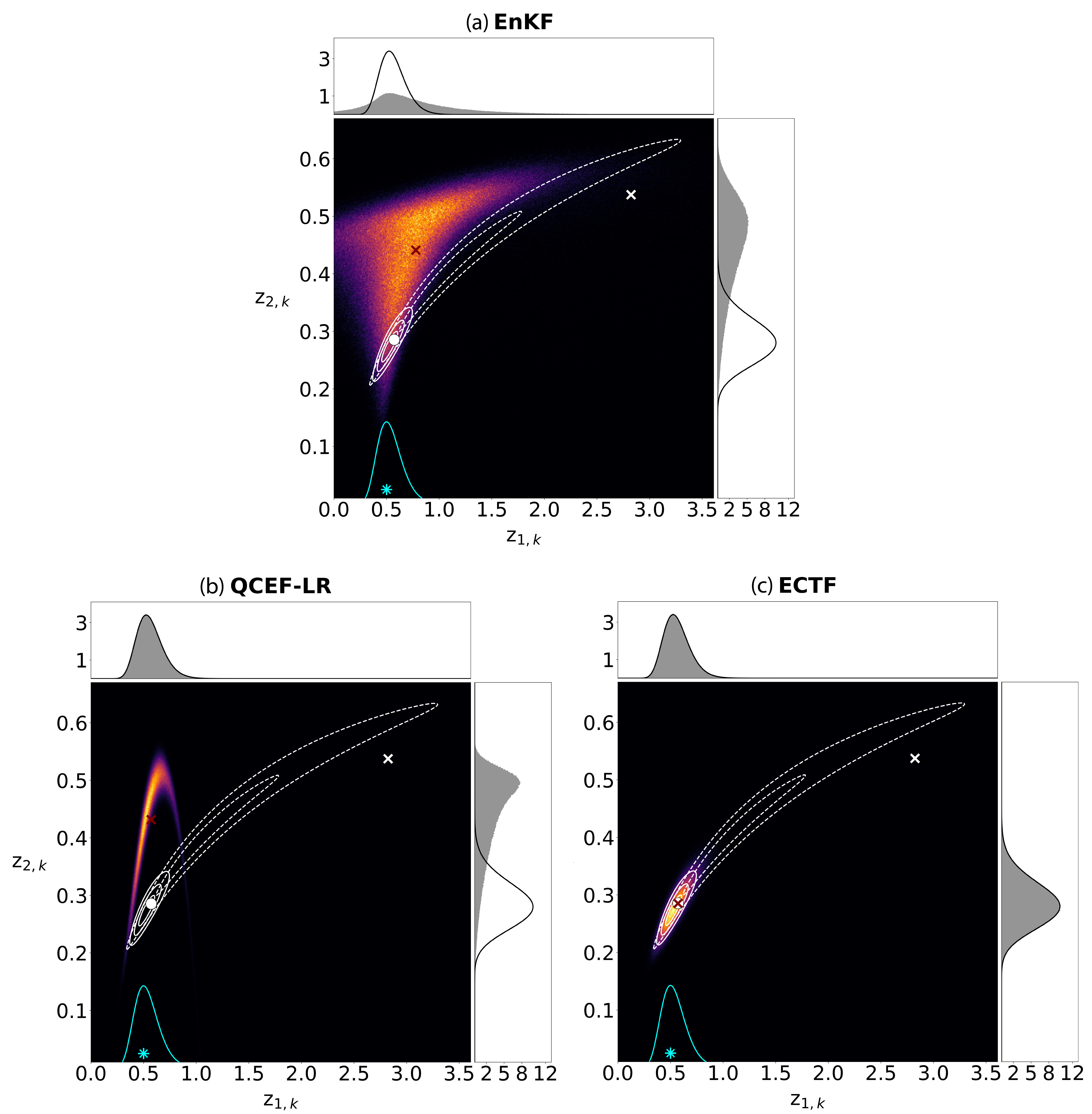}
  \caption{An example comparison between (a) \texttt{EnKF}, (b) \texttt{QCEF-LR} and (c) \texttt{ECTF} in the case of strong prior correlations and accurate observations. Dashed and solid white contours represent the prior and posterior pdfs, whereas the white circle -- the true posterior mean. The likelihood is shown as a cyan curve at the bottom of each panel, with the observation $y_k=0.5$ plotted as a cyan asterisk. The analysis ensemble produced by each filter is color shaded. The white and maroon crosses indicate the prior and posterior ensemble means.}
  \label{fig:example}
\end{figure}

Finally, the ensemble filtering differences are illustrated with a representative numerical example. In Fig.~\ref{fig:example}, the \texttt{EnKF}, \texttt{QCEF-LR} and \texttt{ECTF} updates are compared in the high state correlation/low observation error regime discussed so far, but a value of $r=0.05$ is chosen to enhance visual clarity\footnote{The smaller $r$ is, the more compact the posterior pdf becomes. In the limit $r \rightarrow 0$, the posterior is a delta distribution centered on the observation value $y_k$ (\textit{cf}.~\hyperlink{prop:CTF_obSpace_consistency}{Proposition}), which is hard to visualize.}. It is evident that the prior state variables exhibit a strong nonlinear correlation (dashed white contours), whereas the likelihood is characterized by a sharp peak over the observation $y_k=0.5$. In agreement with \texttt{CTF} theory, the posterior solution (solid white contours) preserves the statistical relationship between the two state variables due to the conjugate nature of the Bayesian update. Moreover, the posterior pdf is drawn closer to the observed value.

As expected from Fig.~\ref{fig:skill_metrics}, the \texttt{EnKF} analysis depicted in Fig.~\ref{fig:example}a is highly biased and overdispersed, with over $7\%$ of the analysis members located outside the physical bounds. Moreover, its Gaussian update appears to introduce two distinct regions: one where the analysis members are uncorrelated ($\rho<0.4$) and another one where they are highly correlated ($\rho>0.4$). 

The statistical relationship between the two state variables is made even worse in \texttt{QCEF-LR} (Fig.~\ref{fig:example}b), which produces a non-monotonic dependence in the analysis solution. In addition, the majority of the analysis members are located away from the posterior pdf. The low skill of \texttt{QCEF-LR} in this example occurs despite the ability of this filter to provide a perfect update in observation space (see top marginal histogram). In fact, comparing only the marginal posteriors of \texttt{EnKF} and \texttt{QCEF-LR} may lead to the deceptive conclusion that the latter method is more accurate. This misleading fact highlights the importance of assessing the multivariate characteristics of non-Gaussian filters -- a point which was also stressed in the concluding remarks of \citet{poterjoy_2022}.  

Finally, Fig.~\ref{fig:example}c shows that \texttt{ECTF} does not only provide consistent marginal updates, but also ensures that the analysis histogram matches the true Bayesian posterior. In addition, the statistical relationship between $\tz_{1,k}$ and $\tz_{2,k}$ is preserved following the analysis step, which is in line with the conjugacy property discussed in Section \ref{subsec:CTF_properties}.

\section{Conclusions}
\label{sec:conclusions}

\vspace*{-1mm}
To summarize, this paper introduced a new nonlinear filtering theory that generalizes the Kalman filter to arbitrarily non-Gaussian distributions while preserving the analytical tractability of the prediction and update steps. The new estimation framework was derived from an alternative state-space model in which linear-Gaussian dynamics are corrected by invertible nonlinear functions (diffeomorphisms). A key feature of the resulting Conjugate Transform Filter (\texttt{CTF}) is that the prior and posterior belong to the same distribution class. This has the practical benefit of preserving the statistical relationships between state variables following the update step. Additional theoretical analysis revealed that a special formulation of \texttt{CTF} converges to the observations when their errors become negligible. 

The closed-form expressions characterizing the new \texttt{CTF} theory were also used to formulate an ensemble filtering approximation (\texttt{ECTF}) suitable for large Earth system models. The key advantage of \texttt{ECTF} is that it can leverage existing DA algorithms: after choosing (or estimating) appropriate nonlinear state and observation functions, an \texttt{EnKF} solver is evoked to update the parameters associated with the latent Gaussian space. The idealized statistical experiments in Section \ref{sec:exps} validated the Bayesian consistency of \texttt{ECTF} and demonstrated the advantages of its multivariate non-Gaussian update. Relative to a benchmark \texttt{EnKF}, the greatest benefits from \texttt{ECTF} showed up when observations errors are small and the prior state is highly correlated. In this regime, \texttt{ECTF} also yielded substantial advantages over a two-step \texttt{QCEF-LR} method, which follows \citet{anderson_2022} to provide an exact non-Gaussian update in observation space, but uses a simple linear regression for correcting the unobserved state variables.

This paper is expected to serve as a foundation for a series of upcoming studies dedicated to further developing the \texttt{ECTF} algorithm. While the idealized experiments presented here revealed that the application of fixed nonlinearities allows \texttt{ECTF} to respect the physical bounds of the modeled system, fixed transformations assume perfect knowledge of the prior uncertainties. Future research will therefore explore the introduction of learnable parameters $\tth_k$ in $\bvf_k$ and $\bvg_k$, which can be estimated directly from the prior ensemble (\textit{cf}.~Section \ref{subsec:ECTF_prior_fitting}). There are many options to parameterize these nonlinear functions, ranging from simple elementwise transformations to much more complex Invertible Neural Networks (INNs), which can fit arbitrary distributions. Therefore, a major goal of future investigations would be to determine the optimal balance between DA accuracy and computational overheads. This will be accomplished through a careful examination of different functional representations and estimation strategies, such as the online and offline fitting of the prior pdf parameters. Special attention will be also given to the construction of appropriate localization and inflation techniques with a view of scaling \texttt{ECTF} to high-dimensional models. Once this goal is accomplished, the resulting algorithms are expected to benefit a wide range of geophysical applications, including the high-resolution forecasting of convective storms and the coupled modeling of Earth system components. The ongoing trends toward more pronounced model nonlinearities and higher observation accuracies will be especially beneficial for \texttt{ECTF} in view of its ability to provide substantial benefits in this setting.

\subsubsection{Acknowledgements.} This research was funded by the National Center for Atmospheric Research, which is a major facility sponsored by the National Science Foundation under Cooperative Agreement No.~1852977. Any opinions, findings, and conclusions or recommendations expressed in this publication are those of the author and do not necessarily reflect the views of the National Science Foundation. The author would like to thank Jeffrey Anderson, Mohamad El Gharamti, Ian Grooms, Ricardo Baptista, Youssef Marzouk, Maximilian Ramgraber, Derek Posselt, Elizabeth Satterfield, John Schreck, Chris Snyder, Peter Jan van Leeuwen, Craig Bishop, Steven Fletcher, Eviatar Bach, Jonathan Poterjoy, Man-Yau (Joseph) Chan, Helen Kershaw, Karen Slater, and Ryan McMichael for useful discussions that greatly benefited this manuscript.

\subsubsection{Data availability statement.}
All code used in this study will be made available on GitHub upon completing the peer-review process for this article.

\vspace*{5mm}
\appendix

\vspace*{2mm}
\begin{center} \large
    \textsc{Appendix}
\end{center}

\vspace*{-4mm}
\section{Notation} \label{appendix:notation}
To better highlight the probabilistic nature of estimation theory, the mathematical symbols used in this paper differ slightly from the standard DA notation introduced by \citet{ide_et_al_1997}. Instead of using bold lowercase letters for the state and observation vectors, deterministic and random versions of these variables are written as regular lowercase and capital letters, respectively. For example, $\ty_k$ denotes a specific realization of the observation process $\{ \tY_k \}_{k \in \mathbb{Z}_{>0}}$ at time $k$. Bold lowercase letters are reserved for vector-valued functions, such as $\bvf_k$ and $\bvg_k$ in (\ref{eq:SSM}), whereas bold capital letters represent matrices (e.g., the observation operator $\bH_k$). Italicized letters collect all scalar objects, examples of which include the Gaussian density $\phi$, the determinant $det$ as well as the parameters $r$ and $\rho$ from Section \ref{sec:exps}.

\section{Proofs}

\subsection{Lemma on closed-form products of non-Gaussian densities} \label{appendix:lemma}

This lemma consists of two parts which give closed-form expressions for products of the non-Gaussian pdfs $p(\tv|\tu)$ and $p(\tu|\tw)$. Notably, the presence of Gaussian base densities makes it possible to adapt the standard technique of completing squares in high dimensions to arrive at the necessary results. Some of the algebraic manipulations can be found in classical textbooks, but they are still presented in sufficient detail here to make this article self-contained.

\subsubsection{Part I: Equations (\ref{eq:lemma_i-pdf})--(\ref{eq:lemma_i-mu}).} In view of (\ref{eq:change_of_variables}), the two conditional pdfs can be explicitly written as

\vspace*{-1mm}
\begin{align}
    & p(\tv|\tu) = \frac{| det\ \bJ_{\bvt_v^{-1}}(\tv) |}{(2 \pi)^{N_v/2} (det\ \bE_v)^{1/2} }\ exp \left(-\frac{1}{2} ||\Ttv - (\bA \Ttu + \tb) ||^2_{\bE_v} \right) \label{eq:condPdf_vu}\\
    & p(\tu|\tw) = \frac{| det\ \bJ_{\bvt_u^{-1}}(\tu) |}{(2 \pi)^{N_u/2} (det\ \bE_u)^{1/2} }\ exp \left(-\frac{1}{2} ||\Ttu - \upmu_u ||^2_{\bE_u} \right) \label{eq:condPdf_uw},
\end{align}
\vspace*{2mm}

\noindent with $||\tx||^2_\bE \coloneqq \tx^\top \bE^{-1} \tx$. Using the quadratic identity 

\vspace*{0.1mm}
\begin{equation} \label{eq:quad_identity}
    ||\tx \pm \ty||^2_\bE = \tx^\top \bE^{-1} \ty \pm 2 \ty^\top \bE^{-1} \tx + \ty^\top \bE^{-1} \ty
\end{equation}
\vspace*{0.1mm}

\noindent and only considering terms proportional to $\tu$ (and hence $\Ttu$), the product $p(\tv|\tu) p(\tu|\tw)$ simplifies to 

\begin{equation} \label{eq:pdf_prod_simple}
    p(\tv|\tu) p(\tu|\tw) \underset{\tu}{\propto} |det\ \bJ_{\bvt_u^{-1}}(\tu)|\ exp \left( -\frac{1}{2} \left( u_1+u_2 \right) \right),
\end{equation}

\noindent where the scalar expressions $u_1,u_2 \in \mathbb{R}$ are given by 

\vspace*{-2mm}
\begin{align}
    & u_1 = \Ttu^\top \bE_u^{-1}  \Ttu - 2\upmu_u^\top \bE_u^{-1} \Ttu + \upmu_u^\top \bE_u^{-1} \upmu_u\\
    & u_2 = \Ttv^\top \bE_v^{-1} \Ttv -2 (\bA \Ttu + \tb)^\top \bE_v^{-1} \Ttv + (\bA \Ttu + \tb)^\top \bE_v^{-1} (\bA \Ttu + \tb) \label{eq:v_scalar}.
\end{align}

If $a \in \mathbb{R}$, it holds that $a = a^\top$. Hence, $\tx^\top \bE^{-1} \ty = \ty^\top \bE^{-1} \tx$ for any $\tx,\ty \in \mathbb{R}^N$ and a precision matrix\footnote{A precision matrix is the inverse of a covariance matrix and is symmetric.} $\bE^{-1} \in \mathbb{R}^{N \times N}$. Based on this fact, the last two terms in (\ref{eq:v_scalar}) can be expanded as $-2 \Ttv^\top \bE_v^{-1} \bA \Ttu - 2\Ttv^\top \bE_v^{-1} \tb$ and $\Ttu^\top \bA^\top \bE_v^{-1} \bA \Ttu + 2\tb^\top \bE_v^{-1} \bA \Ttu + \tb^\top \bE_v^{-1} \tb$, respectively.

Grouping terms according to powers of $\Ttu$, the exponent of (\ref{eq:pdf_prod_simple}) can be rewritten as

\begin{equation} \label{eq:exponent}
    -\frac{1}{2} \left[ \Ttu^\top(\bE_u^{-1}+\bA^\top\bE_v^{-1}\bA)\Ttu -2(\upmu_u^\top\bE_u^{-1}+\Ttv^\top\bE_v^{-1}\bA-\tb^\top\bE_v^{-1}\bA)\Ttu + c_1 \right],
\end{equation}
\vspace*{1mm}

\noindent where $c_1 = \upmu_u^\top\bE_u^{-1}\upmu_u + \Ttv^\top\bE_v^{-1}\Ttv - 2\Ttv^\top\bE_v^{-1}\tb+\tb^\top\bE_v^{-1}\tb = \upmu_u^\top\bE_u^{-1}\upmu_u + ||\Ttv - \tb||^2_{\bE_v}$ collects all terms not dependent on $\Ttu$.

Next, it is desirable to rewrite (\ref{eq:exponent}) in a more compact form such that it resembles the exponents in (\ref{eq:condPdf_vu}) and (\ref{eq:condPdf_uw}). From the quadratic identity (\ref{eq:quad_identity}) with $\tx \leftarrow \Ttu, \ty \leftarrow \upmu_p$ and $\bE \leftarrow \bE_p$, it is clear that

\vspace*{-4mm}
\begin{align}
    & \nonumber \bE_p^{-1} = \bE_u^{-1}+\bA^\top\bE_v^{-1}\bA \\
    \Leftrightarrow\ & \bE_p = \left( \bE_u^{-1}+\bA^\top\bE_v^{-1}\bA \right)^{-1} \label{eq:E_p}.
\end{align}

\noindent For later use, this expression will be rewritten using the Sherman-Morrison-Woodbury (SMW) identity,

\vspace*{-4mm}
\begin{align} \nonumber
    \bE_p &= \bE_u - \bE_u \bA^\top(\bE_v+\bA\bE_u\bA^\top)^{-1}\bA\bE_u \\
    &= (\bI - \bB \bA)\bE_u \label{eq:Pa},
\end{align}

\noindent where 

\vspace*{-4mm}
\begin{align}
    \bB &\coloneqq \bE_u \bA^\top\underbrace{(\bE_v+\bA\bE_u\bA^\top)^{-1}}_{\text{apply SMW identity}} \label{eq:Kgain_long}\\\nonumber
    &= \bE_u \bA^\top [\bE_v^{-1}-\bE_v^{-1}\bA \underbrace{(\bE_u^{-1}+\bA^\top\bE_v^{-1}\bA)^{-1}}_{=\bE_p}\bA^\top\bE_v^{-1}]=\\\nonumber
    &= [\bE_u - \bE_u \bA^\top \bE_v^{-1}\bA \bE_p] \bA^\top \bE_v^{-1}\\\nonumber
    &= [\bE_u \bE_p^{-1} - \bE_u \bA^\top \bE_v^{-1}\bA] \bE_p\bA^\top \bE_v^{-1}\\\nonumber
    &= (\bI + \cancel{\bE_u\bA^\top\bE_v^{-1}\bA}-\cancel{\bE_u \bA^\top \bE_v^{-1}\bA})\bE_p\bA^\top \bE_v^{-1}\\
    &= \bE_p\bA^\top \bE_v^{-1} \label{eq:Kgain_short}.
\end{align}

To find $\upmu_p$, the coefficients associated with the linear terms in (\ref{eq:quad_identity}) and (\ref{eq:exponent}) need to be matched,

\vspace*{-4mm}
\begin{align}
    & \nonumber -2 \upmu_p^\top \bE_p^{-1} = -2(\upmu_u^\top\bE_u^{-1}+\Ttv^\top\bE_v^{-1}\bA-\tb^\top\bE_v^{-1}\bA) \\
    \Leftrightarrow &\ \upmu_p = \bE_p \left[\bE_u^{-1}\upmu_u + \bA^\top \bE_v^{-1} (\Ttv-\tb) \right] \label{eq:mu_p}.
\end{align}

\noindent With a hindsight for subsequent calculations, the above expression for $\upmu_p$ can be represented in an alternative form using (\ref{eq:Pa}) and (\ref{eq:Kgain_short}),

\vspace*{-4mm}
\begin{align} \nonumber
    \upmu_p &= (\bI - \bB \bA) \upmu_u + \bB(\Ttv-\tb)\\
    &= \upmu_u + \bB(\Ttv-\tb-\bA\upmu_u) \label{eq:xa}.
\end{align}

Putting everything together, (\ref{eq:pdf_prod_simple}) becomes

\begin{equation}
    p(\tv|\tu) p(\tu|\tw) \underset{\tu}{\propto} |det\ \bJ_{\bvt_u^{-1}}(\tu)|\ exp \left( -\frac{1}{2} ||\Ttu - \upmu_p ||^2_{\bE_p} \right) \underset{\tu}{\propto} \bvt_{u\, \sharp} \, \phi \left[\tu ;\upmu_p,\bE_p \right],
\end{equation}
\vspace*{2mm}

\noindent where $\upmu_p$ is given either by (\ref{eq:mu_p}) or (\ref{eq:xa}), and $\bE_p$ -- by (\ref{eq:E_p}) or (\ref{eq:Pa}).

$\qed$

\subsubsection{Part II: Equations (\ref{eq:lemma_ii-pdf})--(\ref{eq:lemma_ii-mu}).} The assumption for $\tV$'s conditional independence is first utilized to show the first equality in (\ref{eq:lemma_ii-pdf}),

\vspace*{1mm}
\begin{equation} \label{eq:integral}
    p(\tv|\tw) = \int p(\tv,\tu|\tw) d\tu = \int  p(\tu|\tw) p(\tv|\tu,\tw) d\tu = \int  p(\tv|\tu) p(\tu|\tw) d\tu.
\end{equation}
\vspace*{0.1mm}

\noindent The integrand in the last expression of (\ref{eq:integral}) can be further expanded using (\ref{eq:condPdf_vu}) and (\ref{eq:condPdf_uw}),

\begin{equation} \label{eq:condPdf_vw}
    p(\tv|\tw) = \frac{| det\ \bJ_{\bvt_v^{-1}}(\tv) |}{(2 \pi)^{N_v/2} (2 \pi)^{N_u/2} (det\ \bE_v)^{1/2} (det\ \bE_u)^{1/2}} \int | det\ \bJ_{\bvt_u^{-1}}(\tu) |\ exp \left( -\frac{1}{2} \left( u_1+u_2 \right) \right) d\tu.
\end{equation}
\vspace*{1mm}

Similar to Part I, the subsequent analysis focuses on the exponent $-\frac{1}{2} (u_1+u_2)$ and aims to recover the term $||\Ttu - \upmu_p ||^2_{\bE_p}$. However, the following calculations are slightly more involved as one needs to also consider the terms not depending on $\Ttu$ (or $\tu$). In particular, the free term in (\ref{eq:quad_identity}) can be expanded as

\vspace*{-2mm}
\begin{align} \nonumber
    \upmu_p^\top \bE_p^{-1} \upmu_p &= \left[\bE_u^{-1}\upmu_u + \bA^\top \bE_v^{-1} (\Ttv-\tb) \right]^\top \bE_p^\top  \cancel{\bE_p^{-1}} \cancel{\bE_p} \left[\bE_u^{-1}\upmu_u + \bA^\top \bE_v^{-1} (\Ttv-\tb) \right]\\
    &= || \bE_u^{-1}\upmu_u + \bA^\top \bE_v^{-1} (\Ttv-\tb) ||^2_{\bE_p^{-1}},
\end{align}

\noindent where the symmetry of covariance matrices was used to replace $\bE_p^\top$ with $\bE_p$. To complete the square in $\Ttu$, this free term needs to be added and subtracted from (\ref{eq:exponent}), resulting in 

\begin{equation} 
    exp \left( -\frac{1}{2} \left( u_1+u_2 \right) \right) = exp \left( -\frac{1}{2} \left( ||\Ttu - \upmu_p ||^2_{\bE_p}+c_2 \right) \right),
\end{equation}

\noindent with 

\vspace*{-4mm}
\begin{align} \nonumber
   c_2 &= -|| \bE_u^{-1}\upmu_u + \bA^\top \bE_v^{-1} (\Ttv-\tb) ||^2_{\bE_p^{-1}} + c_1\\
   &= -|| \bE_u^{-1}\upmu_u + \bA^\top \bE_v^{-1} (\Ttv-\tb) ||^2_{\bE_p^{-1}} + \upmu_u^\top\bE_u^{-1}\upmu_u + ||\Ttv - \tb||^2_{\bE_v} \label{eq:c2}.
\end{align}

\noindent Hence, (\ref{eq:condPdf_vw}) transforms to

\begin{align} \nonumber
    p(\tv|\tw) &= \frac{exp(-c_2/2)\ | det\ \bJ_{\bvt_v^{-1}}(\tv) |}{(2 \pi)^{N_v/2} (2 \pi)^{N_u/2} (det\ \bE_v)^{1/2} (det\ \bE_u)^{1/2}} \int | det\ \bJ_{\bvt_u^{-1}}(\tu) |\  exp \left( -\frac{1}{2} \left( ||\Ttu - \upmu_p ||^2_{\bE_p} \right) \right) d\tu \\
    &= \frac{\cancel{(2 \pi)^{N_u/2}} (det\ \bE_p)^{1/2} }{(2 \pi)^{N_v/2} \cancel{(2 \pi)^{N_u/2}} (det\ \bE_v)^{1/2} (det\ \bE_u)^{1/2} }\ exp(-c_2/2)\ | det\ \bJ_{\bvt_v^{-1}}(\tv) |
    \label{eq:condPdf_vw-v2}.
\end{align}
\vspace*{1mm}

\noindent In order to obtain (\ref{eq:condPdf_vw-v2}), the preceeding line was multiplied and divided by $(2 \pi)^{N_u/2} (det\ \bE_p)^{1/2}$ such that

$$
p(\tu) = \frac{| det\ \bJ_{\bvt_u^{-1}}(\tu) |}{(2 \pi)^{N_u/2} (det\ \bE_p)^{1/2}} exp \left( -\frac{1}{2} \left( ||\Ttu - \upmu_p ||^2_{\bE_p} \right) \right)
$$
\vspace*{2mm}

\noindent becomes a valid pdf which integrates to 1.

To proceed further, the exponent in (\ref{eq:condPdf_vw-v2}) also needs to be expressed as a complete square. Expanding the norms in (\ref{eq:c2}) and grouping terms according to powers of $\Ttv' \coloneqq \Ttv-\tb$,

\vspace*{-3mm}
\begin{equation} \label{eq:c2_new}
    c_2 = \Ttv'^\top (\bE_v^{-1} - \bE_v^{-1}\bA \bE_p \bA^\top\bE_v^{-1})\Ttv' - 2\upmu_u^\top\bE_u^{-1}\bE_p\bA^\top\bE_v^{-1}\Ttv'+\upmu_u^\top (\bE_u^{-1}-\bE_u^{-1}\bE_p\bE_u^{-1})\upmu_u.
\end{equation}

Next, this expression is compared against the quadratic identity (\ref{eq:quad_identity}) with $\tx \leftarrow \Ttv', \ty \leftarrow \upmu_i'$ and $\bE \leftarrow \bE_i$. Matching the quadratic terms gives

\noindent
\begin{align} \nonumber
    &\bE_i^{-1} \overset{^\text{(\ref{eq:E_p})}}{=} \bE_v^{-1} - \bE_v^{-1}\bA (\bE_u^{-1}+\bA^\top\bE_v^{-1}\bA)^{-1} \bA^\top\bE_v^{-1} \overset{^\text{SMW id}}{=} (\bE_v + \bA\bE_u\bA)^{-1}\\
    \Leftrightarrow\ &\bE_i = \bA\bE_u\bA^\top+\bE_v \label{eq:E_i}.
\end{align}

Matching the linear terms,

\noindent
\begin{align} \nonumber
    \upmu_i' &= \bE_i \bE_v^{-1}\bA\bE_p\bE_u^{-1}\upmu_u\\\nonumber
    &\overset{^\text{(\ref{eq:E_i}),(\ref{eq:E_p})}}{=} (\bA\bE_u\bA^\top+\bE_v)\bE_v^{-1}\bA \underbrace{(\bE_u^{-1}+\bA^\top\bE_v^{-1}\bA)^{-1}}_{\text{apply SMW identity}}\bE_u^{-1}\upmu_u\\\nonumber
    &= (\bA\bE_u\bA^\top+\bE_v)\bE_v^{-1}\bA \left[ \bE_u-\bE_u\bA^\top(\bA\bE_u\bA^\top+\bE_v)^{-1}\bA\bE_u) \right] \bE_u^{-1}\upmu_u\\\nonumber
    &= (\bA\bE_u\bA^\top\bE_v^{-1}+\bI) \left[\bA-\bA\bE_u\bA^\top(\bA\bE_u\bA^\top+\bE_v)^{-1}\bA \right] \upmu_u\\\nonumber
    &= (\bA\bE_u\bA^\top\bE_v^{-1}+\bI) \left[\bI-\bA\bE_u\bA^\top(\bA\bE_u\bA^\top+\bE_v)^{-1} \right] \bA\upmu_u\\\nonumber
    &= \{\bI+\bA\bE_u\bA^\top [\bE_v^{-1}-(\bA\bE_u\bA^\top+\bE_v)^{-1}-\bE_v^{-1}\bA\bE_u\bA^\top(\bA\bE_u\bA^\top+\bE_v)^{-1} ] \} \bA\upmu_u\\\nonumber
    &= \{\bI+\bA\bE_u\bA^\top [\bE_v^{-1} - (\bI+\bE_v^{-1}\bA\bE_u\bA^\top)(\bA\bE_u\bA^\top+\bE_v)^{-1} ] \} \bA\upmu_u\\\nonumber
    &= \{\bI+\bA\bE_u\bA^\top [\cancel{\bE_v^{-1}} - \cancel{\bE_v^{-1}} \underbrace{(\bE_v+\bA\bE_u\bA^\top)(\bA\bE_u\bA^\top+\bE_v)^{-1}}_{=\bI} ] \} \bA\upmu_u\\
    &= \bA\upmu_u \label{eq:mu_i}.
\end{align}

To see that (\ref{eq:c2_new}) is a complete square, it remains to prove that all remaining terms (which do not depend on $\Ttv'$) also match; i.e., $\upmu_i'^\top \bE_i^{-1} \upmu_i' = \upmu_u^\top (\bE_u^{-1}-\bE_u^{-1}\bE_p\bE_u^{-1})\upmu_u$. This can be done by simplifying the right-hand side expression,

\vspace*{-4mm}
\begin{align} \nonumber
    \upmu_u^\top (\bE_u^{-1}-\bE_u^{-1}\bE_p\bE_u^{-1})\upmu_u &\overset{^\text{(\ref{eq:E_p})}}{=} \upmu_u^\top [\bE_u^{-1}-\bE_u^{-1}\underbrace{(\bE_u^{-1}+\bA^\top\bE_v^{-1}\bA)^{-1}}_{\text{apply SMW identity}}\bE_u^{-1}]\upmu_u\\\nonumber
    &= \upmu_u^\top \{ \bE_u^{-1}-\bE_u^{-1}[\bE_u-\bE_u\bA^\top(\bA\bE_u\bA^\top+\bE_v)^{-1}\bA\bE_u]\bE_u^{-1}  \} \upmu_u\\\nonumber
    &= \upmu_u^\top \{ \cancel{\bE_u^{-1}}-\cancel{\bE_u^{-1}}+\bA^\top(\bA\bE_u\bA^\top+\bE_v)^{-1}\bA \} \upmu_u\\\nonumber
    &= \upmu_u^\top \bA^\top(\bA\bE_u\bA^\top+\bE_v)^{-1}\bA \upmu_u\\\nonumber
    &\overset{^\text{(\ref{eq:E_i}),(\ref{eq:mu_i})}}{=} \upmu_i'^\top \bE_i^{-1} \upmu_i'.
\end{align}

\noindent Hence, (\ref{eq:c2_new}) becomes

\vspace*{0.1mm}
\begin{equation} \label{eq:c2_final}
    c_2 = ||\Ttv'-\upmu_i'||^2_{\bE_i} = ||\Ttv-\tb-\bA\upmu_u||^2_{\bE_i} = ||\Ttv-\upmu_i||^2_{\bE_i},
\end{equation}
\vspace*{0.1mm}

\noindent with $\upmu_i \coloneqq \bA\upmu_u + \tb$.

Inserting the above expression for $c_2$ back in (\ref{eq:condPdf_vw-v2}) gives 

\vspace*{-2mm}
\begin{align} \nonumber
    p(\tv|\tw) &= \frac{|det\ \bJ_{\bvt_v^{-1}}(\tv)|}{(2 \pi)^{N_v/2} (\frac{det\ \bE_v\ det\ \bE_u}{det\ \bE_p})^{1/2} }\ exp\left(-\frac{1}{2} ||\Ttv-\upmu_i||^2_{\bE_i}\right)\\
    &= \frac{| det\ \bJ_{\bvt_v^{-1}}(\tv) |}{(2 \pi)^{N_v/2} (det\ \bE_i)^{1/2} }\ exp \left(-\frac{1}{2} ||\Ttv - \upmu_i ||^2_{\bE_i} \right) =: \bvt_{v\, \sharp} \, \phi \left[\tv ;\upmu_i,\bE_i \right].
\end{align}
\vspace*{2mm}

\noindent Note that simplification of the determinants in the second line follows from (4.39) in \citet{cohn_1997}. Specifically, (4.39) is first rearranged to get $det\ \bM = (det\ \bR)(det\ \bP^f)(det\ \bP^a)^{-1}$ and then the following substitutions need to be made: $\bM \leftarrow \bE_i$, $\bR \leftarrow \bE_v$, $\bP^f \leftarrow \bE_u$, $\bP^a \leftarrow \bE_p$ and $\bH \leftarrow \bA$. 

$\qed$

\subsection{Theorem on exact nonlinear filtering} \label{appendix:theorem}

The proof is inductive and follows the strategy outlined in Appendix A.1 of \citet{de_bezenac_et_al_2020}.

\subsubsection{1) Base case:} \textit{Prove that $p(\tx_1 | \ty_1) = \bvf_{\tth_1\, \sharp} \, \phi \left[\tx_1 ;\upmu_{1|1},\bE_{1|1} \right]$ with $\upmu_{1|1} = \upmu_{1|0}+\bK_1 \left( \widetilde \ty_1 - \bH_1 \upmu_{1|0} \right)$, $\bE_{1|1} = \left( \bI-\bK_1 \bH_1  \right) \bE_{1|0}$ and $\bK_1 = \bE_{1|0} \bH_1^\top \left( \bH_1 \bE_{1|0} \bH_1^\top + \bR_1  \right)^{-1}$.}

Applying the conditional pdf definition twice, 

\vspace*{1mm}
\begin{equation} \label{eq:p_x1y1}
    p(\tx_1|\ty_1) = \frac{p(\tx_1,\ty_1)}{p(\ty_1)} \underset{\tx_1}{\propto} p(\tx_1)p(\ty_1|\tx_1).
\end{equation}
\vspace*{1mm}

From (\ref{eq:cond_pdfs-new_SSM-ob}) with $k=1$, $p(\ty_1|\tx_1) = \bvg_{\tth_1 \, \sharp} \, \phi \left[\ty_1; \bH_1 \widetilde \tx_1, \bR_1  \right]$. Relating $p(\tx_1)$ to known quantities requires a marginalization over the joint pdf of $\tX_0$ and $\tX_1$,

\vspace*{1mm}
\begin{equation}
    p(\tx_1) = \int p(\tx_0,\tx_1) d\tx_0 = \int p(\tx_1|\tx_0) p(\tx_0) d\tx_0,
\end{equation}
\vspace*{1mm}

\noindent where $p(\tx_0) = \bvf_{\tth_0 \, \sharp} \, \phi \left[\tx_0 ;\upmu_0,\bE_0 \right]$ using assumption (\ref{eq:initial_pdf}) and $p(\tx_1|\tx_0) = \bvf_{\tth_1 \, \sharp} \, \phi \left[\tx_1; \bM_1 \widetilde \tx_0, \bQ_0  \right]$ from (\ref{eq:cond_pdfs-new_SSM-state}) with $k=1$. 

Applying Part II of Lemma \ref{appendix:lemma} with $p(\tu|\tw) \leftarrow p(\tx_0)$, $p(\tv|\tu) \leftarrow p(\tx_1 | \tx_0)$ and $\tb=0$ gives

\vspace*{-4mm}
\begin{align}
    &p(\tx_1) = \bvf_{\tth_1 \, \sharp} \, \phi \left[\tx_1 ;\upmu_{1|0},\bE_{1|0} \right] \label{eq:p_x1},\\
    &\upmu_{1|0} = \bM_1 \upmu_0,\\
    &\bE_{1|0} = \bM_1 \bE_0 \bM_1^\top + \bQ_0.
\end{align}

\noindent Inserting (\ref{eq:p_x1}) into (\ref{eq:p_x1y1}) and using Part I of Lemma \ref{appendix:lemma} (with the same substitutions as Part II) yields

\vspace*{-4mm}
\begin{align}
    &p(\tx_1|\ty_1) \underset{\tx_1}{\propto} \bvf_{\tth_1 \, \sharp} \, \phi \left[\tx_1 ;\upmu_{1|1},\bE_{1|1} \right] \label{eq:p_x1y1_v2},\\
    &\bE_{1|1} = (\bE_{1|0}^{-1}+\bH_1^\top\bR_1^{-1}\bH_1)^{-1}  \label{eq:E_11},\\
    &\upmu_{1|1} = \bE_{1|1} \left[ \bE_{1|0}^{-1}\upmu_{1|0} + \bH_1^\top \bR_1^{-1} \widetilde \ty_1 \right].
\end{align}
\vspace*{-1mm}

\noindent Note that $p(\tx_1|\ty_1)$ is a valid pdf in view of the normalizing constant $p(\ty_1)$ in (\ref{eq:p_x1y1}), making it possible to drop the proportionality symbol in (\ref{eq:p_x1y1_v2}). Finally, the desired expressions for $\upmu_{1|1}$ and $\bE_{1|1}$ can be obtained from (\ref{eq:Pa}), (\ref{eq:Kgain_long}) and (\ref{eq:xa}).

$\qed$ 

\vspace*{-1.5mm}
\subsubsection{2) Inductive step:} \textit{Assume}

\begin{equation} \label{eq:induc_assump}
    p(\tx_{k-1} | \ty_{k-1}) = \bvf_{\tth_{k-1}\, \sharp} \, \phi \left[\tx_{k-1} ;\upmu_{k-1|k-1},\bE_{k-1|k-1} \right]
\end{equation}
\vspace*{1mm}

\noindent \textit{and prove that $p(\tx_{k} | \ty_{k}) = \bvf_{\tth_{k}\, \sharp} \, \phi \left[\tx_{k} ;\upmu_{k|k},\bE_{k|k} \right]$ with the mean and covariance parameters of the base Gaussian pdf given by (\ref{eq:CTF_theorem-mu})-(\ref{eq:CTF_theorem-gain}).}

The calculations here are nearly identical to the base case, but involve more complex pdfs. A proof sketch is presented, which begins by recognizing that the object of interest is given by (\ref{eq:filtering_recursions-update}). That is,

\vspace*{1mm}
\begin{equation} \label{eq:p_xky1k}
    p(\tx_k | \ty_{1:k}) = \frac{p(\tx_k|\ty_{1:k-1})p(\ty_k|\tx_k)}{\int p(\tx_k|\ty_{1:k-1})p(\ty_k|\tx_k) d\tx_k}\ \underset{\tx_k}{\propto}\ p(\tx_k|\ty_{1:k-1})p(\ty_k|\tx_k).
\end{equation}
\vspace*{1mm}

\noindent To leverage the inductive assumption (\ref{eq:induc_assump}), $p(\tx_k|\ty_{1:k-1})$ in (\ref{eq:p_xky1k}) is rewritten based on the Chapman-Kolmogorov equation (\ref{eq:filtering_recursions-prediction}). Additionally using (\ref{eq:cond_pdfs-new_SSM}) in the resulting expression and applying Lemma \ref{appendix:lemma} yields

\vspace*{-2mm}
\begin{align} \nonumber
    &p(\tx_k | \ty_{1:k}) \overset{^{(\ref{eq:filtering_recursions-prediction})}}{\underset{\tx_k}{\propto}} p(\ty_k|\tx_k) \int p(\tx_k | \tx_{k-1}) p(\tx_{k-1}|\ty_{1:k-1}) d\tx_{k-1}\\\nonumber
    &\overset{^{(\ref{eq:cond_pdfs-new_SSM}),(\ref{eq:induc_assump})}}{=} \bvg_{\tth_k \, \sharp} \, \phi \left[\ty_k; \bH_k \widetilde \tx_k, \bR_k  \right] \int \bvf_{\tth_k \, \sharp} \, \phi \left[\tx_k; \bM_k \widetilde \tx_{k-1}, \bQ_{k-1}  \right] \bvf_{\tth_{k-1}\, \sharp} \, \phi \left[\tx_{k-1} ;\upmu_{k-1|k-1},\bE_{k-1|k-1} \right] d\tx_{k-1}\\\nonumber
    &\overset{^\text{(Lemma \ref{appendix:lemma}, Part II)}}{=} \bvg_{\tth_k \, \sharp} \, \phi \left[\ty_k; \bH_k \widetilde \tx_k, \bR_k  \right]\ \bvf_{\tth_k \, \sharp} \, \phi \left[\tx_k; \upmu_{k|k-1}, \bE_{k|k-1}  \right]\\
    &\overset{^\text{(Lemma \ref{appendix:lemma}, Part I)}}{\underset{\tx_k}{\propto}} \bvf_{\tth_k \, \sharp} \, \phi \left[\tx_k; \upmu_{k|k}, \bE_{k|k}  \right] \label{eq:p_xky1k_final}.
\end{align}
\vspace*{1mm}

\noindent Once again, the proportionality symbol above can be dropped on account of the fact that $p(\tx_k | \ty_{1:k})$ is a valid pdf.

$\qed$

\subsection{Corollary on generalizing the \texttt{KF}} \label{appendix:corollary}

The first part of this corollary follows trivially. However, it is important to emphasize that $\upmu_{k|k}$ in (\ref{eq:CTF_theorem-mu}) and $\bE_{k|k}$ in (\ref{eq:CTF_theorem-sigma}) change their meaning from posterior parameters in the latent Gaussian space to the conditional mean and covariance of the posterior state estimate given by the Kalman filter update.  

For the second part, let $\bvf_{\tth_k}: \mathbb{R}^{N_x} \ni \widetilde \tX \mapsto \bL_x \widetilde \tX + \tb_x \in \mathbb{R}^{N_x}$ and $\bvg_{\tth_k}: \mathbb{R}^{N_y} \ni \widetilde \tY \mapsto \bL_y \widetilde \tY + \tb_y \in \mathbb{R}^{N_y}$ be two affine maps for which $\{\bL_x,\bL_y\}$ and $\{\tb_x,\tb_y\}$ are pairs of invertible matrices and vectors of conforming size. A well known result in probability theory states that Gaussian random vectors are closed under affine transformations. In particular, if $\widetilde \tX \sim \mathcal{N}(\upmu_x,\bE_x) \Rightarrow \tX = \bL_x \widetilde \tX + \tb_x \sim \mathcal{N}(\bL_x \upmu_x + \tb_x,\bL_x \bE_x \bL_x^\top)$. Applying this result to (\ref{eq:CTF_theorem-pdf}) gives

\vspace*{1mm}
\begin{equation} \label{eq:gauss_posterior}
    p(\tx_k | \ty_{1:k}) = \phi \left[\tx_k ;\bL_x \upmu_{k|k} + \tb_x,\bL_x \bE_{k|k} \bL_x^\top \right]\ \ \forall k \in \mathbb{Z}_{>0}.
\end{equation}
\vspace*{-0.5mm}

\noindent Further setting $\tb_x=\tb_y=0$ produces Gaussian parameters identical to those shown in Section 3.5.1 of \citet{snyder_2015}, but without the need to perform lengthy matrix calculations. Note that while the posterior covariance is not explicitly derived in \citet{snyder_2015}, a straightforward calculation shows that

\vspace*{-5mm}
\begin{align} \nonumber
    \bP^a_{zz} &= (\bI-\widetilde \bK \widetilde \bH) \bP^f_{zz}\\\nonumber
    &= (\bL_x \bL_x^{-1}-\bL_x \bK \cancel{\bL_y^{-1}} \cancel{\bL_y} \bH \bL_x^{-1}) \bL_x \bP^f \bL_x^\top\\
    &= \bL_x (\bI-\bK\bH)\cancel{\bL_x^{-1}} \cancel{\bL_x} \bP^f \bL_x^\top = \bL_x \bP^a \bL_x^\top,
\end{align}
\vspace*{1mm}

\noindent which matches the covariance expression in (\ref{eq:gauss_posterior}). 

$\qed$

\vspace*{-2mm}
\subsection{Proposition on the observation-space consistency of \texttt{CTF}} \label{appendix:proposition}

Since the focus here is on the observation-space characteristics of \texttt{CTF}, the first step is to apply the matrix $\bH_k$ to $\TtZ_k | \tY_{1:k}$, the latent Gaussian state at time $k$ conditioned on the history of observations up to that time. Following the same reasoning as in Corollary \ref{appendix:corollary}, it follows that $p(\bH_k \Ttz_k | \ty_{1:k}) = \phi(\bH_k \Ttz_k; \bH_k \upmu_{k|k}, \bH_k \bE_{k|k} \bH_k^\top)$. Using (\ref{eq:CTF_theorem-mu}) together with the assumption that $\bR_k = \bO_{N_y \times N_y}$ gives

\vspace*{1mm}
\begin{equation} \label{eq:H_mu}
    \bH_k \upmu_{k|k} = \cancel{\bH_k \upmu_{k|k-1}} + \underbrace{\bH_k \bE_{k|k-1}\bH_k^\top (\bH_k \bE_{k|k-1} \bH_k^\top + \bO_{N_y \times N_y})^{-1}}_{=\bI} (\Tty_k -\cancel{\bH_k \upmu_{k|k-1}}) = \Tty_k.
\end{equation}
\vspace*{0.1mm}

\noindent Applying the same assumption to (\ref{eq:CTF_theorem-sigma}),

\vspace*{-2mm}
\begin{align} \nonumber
    \bH_k \bE_{k|k} \bH_k^\top &= \bH_k \left[\bI - \bE_{k|k-1}\bH_k^\top(\bH_k\bE_{k|k-1}\bH_k^\top+\bO_{N_y \times N_y})^{-1}\bH_k \right] \bE_{k|k-1} \bH_k^\top \\
    &= \cancel{\bH_k\bE_{k|k-1}\bH_k^\top}-\cancel{\bH_k\bE_{k|k-1}\bH_k^\top} \underbrace{(\bH_k\bE_{k|k-1}\bH_k^\top)^{-1}\bH_k\bE_{k|k-1}\bH_k^\top}_{=\bI} = \bO_{N_y \times N_y} \label{eq:H_E}.
\end{align}

\vspace*{1mm}
\noindent In other words, (\ref{eq:H_mu}) and (\ref{eq:H_E}) imply that in the limit $\bR_k \to \bO_{N_y \times N_y}$,

\vspace*{-2mm}
\begin{align} \nonumber
    &p(\bH_k \Ttz_k | \ty_{1:k}) = \delta(\bH_k \Ttz_k - \Tty_k)\\
    \Leftrightarrow\ \ & \bH_k \TtZ_k | \tY_{1:k} \overset{^\text{a.s.}}{=\joinrel=} \Tty_k = \bvg_{\tth_k}^{-1} (\ty_k). \label{eq:dirac}
\end{align}
\vspace*{-0.5mm}

\noindent Further leveraging the assumption that $\bvg_{\tth_k} = \bvf_{\tth_k^y}$, the expression in (\ref{eq:dirac}) can be modified to yield the desired result,

\vspace*{-4mm}
\begin{align} \nonumber
    \Leftrightarrow\ \ &\widetilde{\bvh_k(\tX_k)} | \tY_{1:k} \overset{^\text{a.s.}}{=\joinrel=} \bvf_{\tth_k^y}^{-1} (\ty_k)\\\nonumber
    \Leftrightarrow\ \ & \bvf_{\tth_k^y} \left( \widetilde{\bvh_k(\tX_k)} \right) | \tY_{1:k} \overset{^\text{a.s.}}{=\joinrel=} \ty_k\\\nonumber
    \Leftrightarrow\ \ & \bH_k \tZ_k | \tY_{1:k} \overset{^\text{a.s.}}{=\joinrel=} \ty_k\\
    \Leftrightarrow\ \ & p \left( \bH_k \tz_k | \ty_{1:k} \right) = \delta(\bH_k \tz_k -\ty_k). 
\end{align}
$\qed$

\vspace*{5mm}
\renewcommand\bibname{\Large{References}}
\bibliographystyle{ametsoc2014}
\bibliography{manuscript}

\end{document}